\documentclass[fleqn,usenatbib]{mnras}
\usepackage[T1]{fontenc}

\usepackage{newtxtext,newtxmath}

\usepackage{ae,aecompl}
\usepackage{graphicx}	% Including figure files
\usepackage{amsmath}	% Advanced maths commands
\usepackage{bm}	% Extra maths symbols
\usepackage{hyperref}
\newcommand{\hMpc}{ h^{-1}{\rm Mpc}}

\newcommand{\hkpc}{ h^{-1}{\rm kpc}}
\newcommand{\ihMpc}{ h\,{\rm Mpc}^{-1}}
\newcommand{\zstar}{\rm z^*}
\title[]{3\%-accurate predictions for the clustering of dark matter, haloes and subhaloes, over a wide range of cosmologies and scales.}
\author[S. Contreras et al.]{
S. Contreras,$^{1}$\thanks{E-mail: sergio.contreras@dipc.org}
R. E. Angulo,$^{1,2}$ M. Zennaro,$^{1}$ G. Aric\`o,$^{1}$ \& M. Pellejero-Iba\~nez$^{1}$.
\\
$^{1}$Donostia International Physics Center (DIPC), Manuel Lardizabal Ibilbidea, 4, 20018 Donostia, Gipuzkoa, Spain.\\
$^{2}$IKERBASQUE, Basque Foundation for Science, 48013, Bilbao, Spain.
}

\date{Accepted XXX. Received YYY; in original form ZZZ}

\pubyear{2020}

\begin{document}
\label{firstpage}
\pagerange{\pageref{firstpage}--\pageref{lastpage}}
\maketitle

\begin{abstract}
Predicting the spatial distribution of objects as a function of cosmology is an essential ingredient for the exploitation of future galaxy surveys. In this paper we show that a specially-designed suite of gravity-only simulations together with cosmology-rescaling algorithms can provide the clustering of dark matter, haloes, and subhaloes with high precision. Specifically, with only 3 $N$-body simulations we obtain the power spectrum of dark matter at $z=0$ and $z=1$ to better than 3\% precision for essentially all currently viable values of 8 cosmological parameters, including massive neutrinos and dynamical dark energy, and over the whole range of scales explored, 0.03 < $k/\ihMpc$ < 5. This precision holds at the same level for mass-selected haloes and for subhaloes selected according to their peak maximum circular velocity. As an initial application of these predictions, we successfully constrain $\Omega_{\rm m}$, $\sigma_8$, and the scatter in subhalo-abundance-matching employing the projected correlation function of mock SDSS galaxies. 

\end{abstract}


%%%%%%%%%%%%%%%%%%%%%%%%%%%%%%%%%%%%%%%%%%%%%%%%%%

%%%%%%%%%%%%%%%%% BODY OF PAPER %%%%%%%%%%%%%%%%%%

\section{Introduction}
\label{sec:Introduction}

With a new generation of galaxy surveys soon to arrive (e.g. Euclid, J-PAS, DESI, LSST, 4MOST), new opportunities for improving our understanding of cosmology and galaxy formation will emerge. For instance, the high accuracy of upcoming galaxy clustering measurements could allow us to distinguish between alternative gravity theories or between cosmological models with or without massive neutrinos.

To take full advantage of these opportunities, we require an accurate modelling of galaxy formation and the dark matter haloes that host them. Although the abundance and spatial distribution of these haloes are, in principle, fully determined by the initial perturbation spectrum and the law of gravity, the detailed way in which galaxies populate them depends on much less understood baryonic processes (star formation, feedback, quenching, etc.). 

The most accurate way to model galaxy clustering as a function of cosmology is through cosmological numerical simulation. For instance, high-resolution dark-matter only simulations enable highly realistic and sophisticated modelling, which faithfully reproduce the abundance, properties, and clustering of observed galaxies. Examples of these models are: abundance matching \citep[SHAM,][]{Conroy:2006,Reddick:2013,Chaves-Montero:2016,Lehmann:2017,Dragomir:2018}, subhalo clustering and abundance matching \citep[SCAM,][]{Guo:2016}, semi-analytical models \citep[SAMs,][]{Henriques:2015,Stevens:2016,Lacey:2016,Croton:2016,Lagos:2018}, and hybrid models like the universe machine \citep{Behroozi:2019} and EMERGE \citep{Moster:2018}. Low-resolution simulations can still model some aspects of the galaxy population albeit via simpler techniques that use only the information of dark matter haloes, e.g halo occupation distribution models \citep[HOD,][]{Peacock:2000, Berlind:2002, Cooray:2002, Zheng:2005}, or density-field based approaches \citep{DeRose:2019b}. 

One way to employ the aforementioned models to predict galaxy clustering as a function of cosmology is to run a large suite of $N$-body simulations with many different cosmologies. Gaussian processes and emulators are typically employed to interpolate between the finite sampling of cosmological parameters. This method has become popular in the last years as it allows a straightforward use of current computational resources (eg. \citealt{Heitmann:2016,AEMULUS1,AEMULUS2, AEMULUS3, AEMULUS4,Nishimichi:2018}). 
The main limitation of this methodology is that, since it is necessary to run typically hundreds of $N$-body simulations, each of these has poor resolution. This restricts the galaxy modelling to the simplest approaches, which in turn increases the uncertainty and possibly even biases cosmological analyses. 

An example of the limitations of simple galaxy modellings such as HODs is that they cannot capture the so-called galaxy assembly bias \citep{Croton:2007}. This is the consequence of two effects: i) the dependence of the halo clustering on properties other than mass \citep{Gao:2005}, and ii) the dependence of the mean occupation of galaxies on halo properties other than mass \citep[a.k.a. occupancy variation][]{Zehavi:2018,Artale:2018,C19}. Although some HOD models attempt to include assembly bias \citep{Hearin:2016,Zehavi:2019}, they are not yet capable of reproducing the full signal. Another limitation of HODs is that they require  modelling of multiple additional effects to obtain accurate predictions: velocity bias \citep{Guo:2015}, satellite abundance distribution \citep{Jimenez:2019}, baryonic effects on halo mass \citep{Beltz-Mohrmann:2019}, or the spatial distribution of galaxies inside haloes \citep{Yuan:2018}. All these add additional parameters and/or complexity to the HOD, making them model dependent and limiting the constraining power of galaxy clustering data, especially on small scales.

An approach that allows the use of high-resolution $N$-body  simulations and thus of realistic galaxy modelling is the "cosmology-rescaling" algorithms \citep{Angulo:2010, Angulo:2015}. This approach avoids the need of running a large number of simulations by rescaling the outputs of a single simulation to predict structure formation in any nearby cosmology. This process is accurate and fast -- taking a few seconds in a normal laptop instead of thousands of CPU-hours that it would take to run a simulation. Although there are potentially more sources of uncertainty, the key difference of this approach is that focus computational resources on a few high-resolution simulations. This allows a better description of non-linear structure and the use of the most sophisticated galaxy formation modelling available. Ultimately, this results in a theoretical description of galaxy clustering that is more physically motivated, accurate, and predictive as it contains less free parameters.

The scaling technique has been validated for multiple summary statistics (dark matter clustering, haloes, halo mass function, weak lensing maps, etc), but always rescaling a simulation with a choice of cosmological parameters set with other motivations (e.g. to be compatible with the latest observational constraints), and scaled to a few target cosmologies. Here, for the first time, we show that the uncertainty of the scaling is predictable, and so it can be used to define an optimal set of cosmological parameters to be rescaled. 

We carry out such optimal simulation suite and perform a systematic study of the accuracy of the cosmology-rescaling technique. For this, we carry out $\sim$ 70  $N$-body simulations varying eight different cosmological parameters:  $\rm \sigma_8$, $\rm \Omega_{\rm m}$, $\rm \Omega_b$, $\rm n_s$, $\rm h$, $M_{\nu}$, $w_0$, and $w_a$ over a range of roughly $10\sigma$ around the best fit parameters of \citealt{Planck13_16}. We quantify the accuracy of the rescaling suite for the power spectrum of dark matter particles, haloes, and subhaloes in both real and redshift space and at $z=0$ and $z=1$. We find that for most cases and scales considered, the precision of the method is better than 3\%. 

As an application of these specially-designed simulations, we test their capability at constraining cosmological parameters from the projected correlation function of SDSS-like galaxies. For this, we perform a Bayesian parameter estimation on mock data and model it using SHAM. We show that our approach is feasible and it delivers tight and unbiased constrains for the parameters we explore.

The layout of the paper is as follows. In \S\ref{sec:Scaling} we introduce the scaling technique, discussing its accuracy and computational requirements. In \S\ref{sec:Opt_Cosmo} we explore the best combinations of cosmologies that maximize the precision of scaling over a given cosmological parameter space. The main results of this work are shown in \S\ref{sec:main} where we quantify the precision of rescaling over a broad range of cosmologies. In \S\ref{sec:MCMC} we test the ability of the scaling to constrain cosmology by modelling the projected correlation function of a SDSS-mock galaxies. We finalize by summarizing our results in \S\ref{sec:Conclusions}.

\section{The Cosmology Scaling technique}
\label{sec:Scaling}

We start this section by recapping the main ingredients and ideas of the cosmology rescaling method (\S\ref{sec:background}). Then, we discuss its limitations in terms of accuracy (\S\ref{sec:halomodel}) and computational requirements (\S\ref{sec:comp_lim}). 

\subsection{The algorithm}
\label{sec:background}

The scaling technique \citep{Angulo:2010} consists in modifying the outputs of a $N$-body simulation that adopts a given (original) cosmology so that it mimics a different (target) cosmology. The specific steps of the algorithm are:

\begin{itemize}
    \item Find the expansion factor and length transformations ($a \rightarrow a^{*}$ and $x \rightarrow x' \equiv s\,x$, respectively) that minimize the difference between the linear mass variance in the original and target cosmologies.
    \item Choose the simulation output with the closest expansion factor to $a^{*}$.
    \item Multiply the positions of simulation particles and objects by $s$ and their masses by $\alpha$, where:
    \begin{equation}
        \alpha \equiv  s^3\Omega_{\rm m}^{\rm target}h_{\rm target}^2/\Omega_{\rm m}^{\rm original}h_{\rm original}^2
    \end{equation}
    \item Subtract the large scale modes from the positions and velocities and add those expected in the target cosmology. This is achieved finding (with a recursive algorithm) the Lagrangian positions in the original simulation and applying to them a 2LPT displacement field that matches the large scale clustering of the target cosmology.
    \item Correct intra-halo velocities to match the virialization condition in the target cosmology as
    \begin{equation}
        \Delta v \rightarrow \Delta v' \equiv \Delta v \, \bigg[  \dfrac{\alpha\ a}{s^3\ a^{*}}   \bigg]^{1/2}
    \end{equation}
\end{itemize}

We refer the reader to \cite{Angulo:2010} for a more comprehensive description of the scaling technique.

In addition to these steps, here we extend the algorithm by correcting the distribution of the particles inside haloes according to the expected differences in concentration-mass relation of haloes in the rescaled and target cosmologies. We refer to this as ``concentration correction'', and it will be explained in detail in \S\ref{sec:Improvements}. Finally, to account for the scale dependence introduced by massive neutrinos on the distribution of matter, we follow the techniques presented by \cite{Zennaro:2017}.

%%%%%%%%%%%%%%%%%%%%%%%%%%%%%%%%%%%%%%%%%%%%%%%%%%%%%%%%%%%%%%%%%%%%%%%%%%%%%%

\subsection{Limitations of the scaling}

The scaling algorithm produces very accurate results for the non-linear mass power spectrum in real and redshift space \citep{Angulo:2010}; the abundance, position, and mass of dark matter haloes \citep{Ruiz:2011}; weak lensing mass profiles \citep{Renneby:2018}, and even in cosmologies with modified gravity \citep{Mead:2015}, with massive neutrinos \citep{Zennaro:2019} or considering the effects of baryons in the mass distribution \citep{Arico:2020}.
However, the algorithm is by no means perfect and also imposes certain requirements to the original simulation. We discuss these aspects next.

\subsubsection{Accuracy}
\label{sec:halomodel}

\begin{figure}
\includegraphics[width=0.45\textwidth]{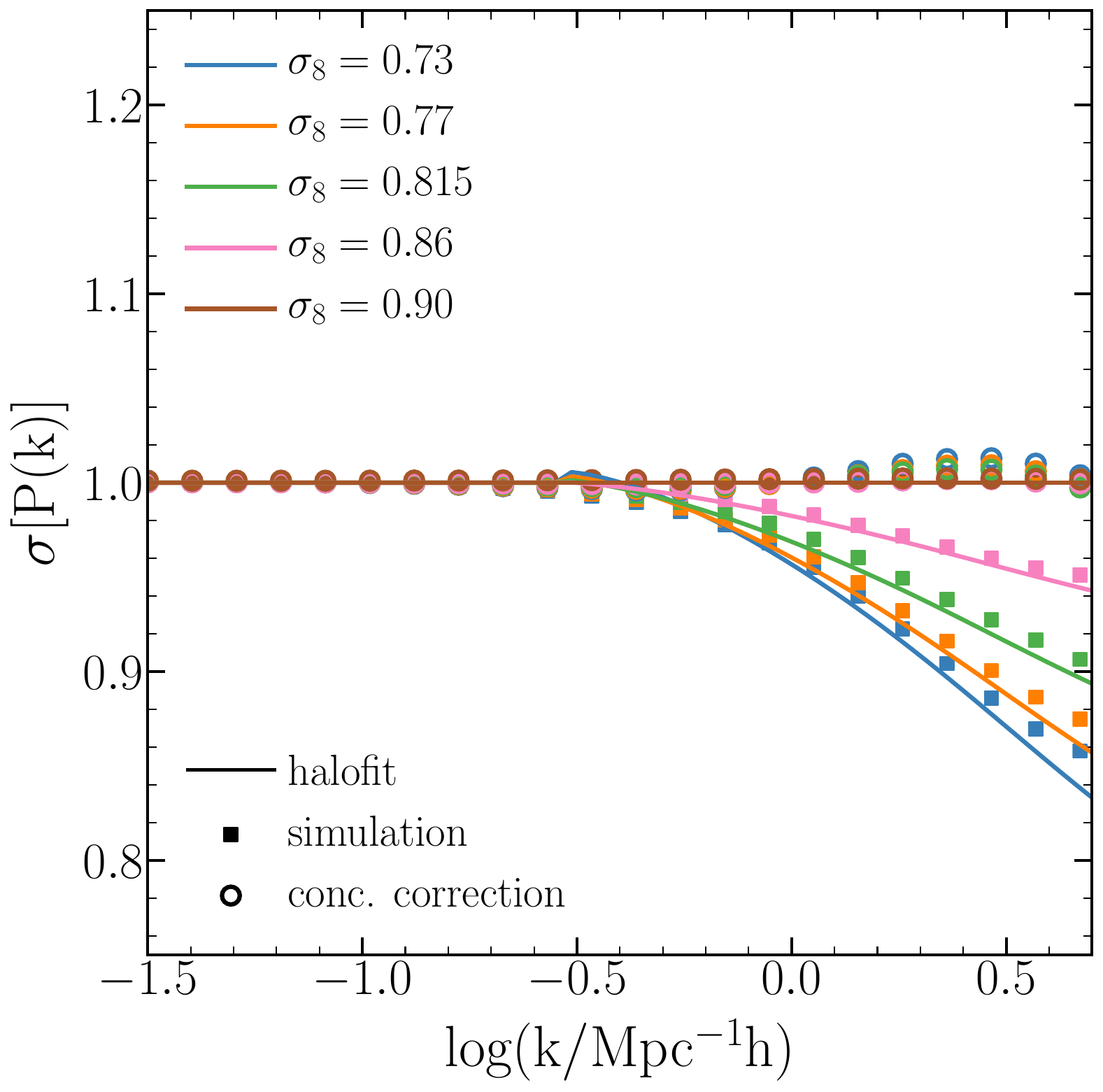}
\includegraphics[width=0.45\textwidth]{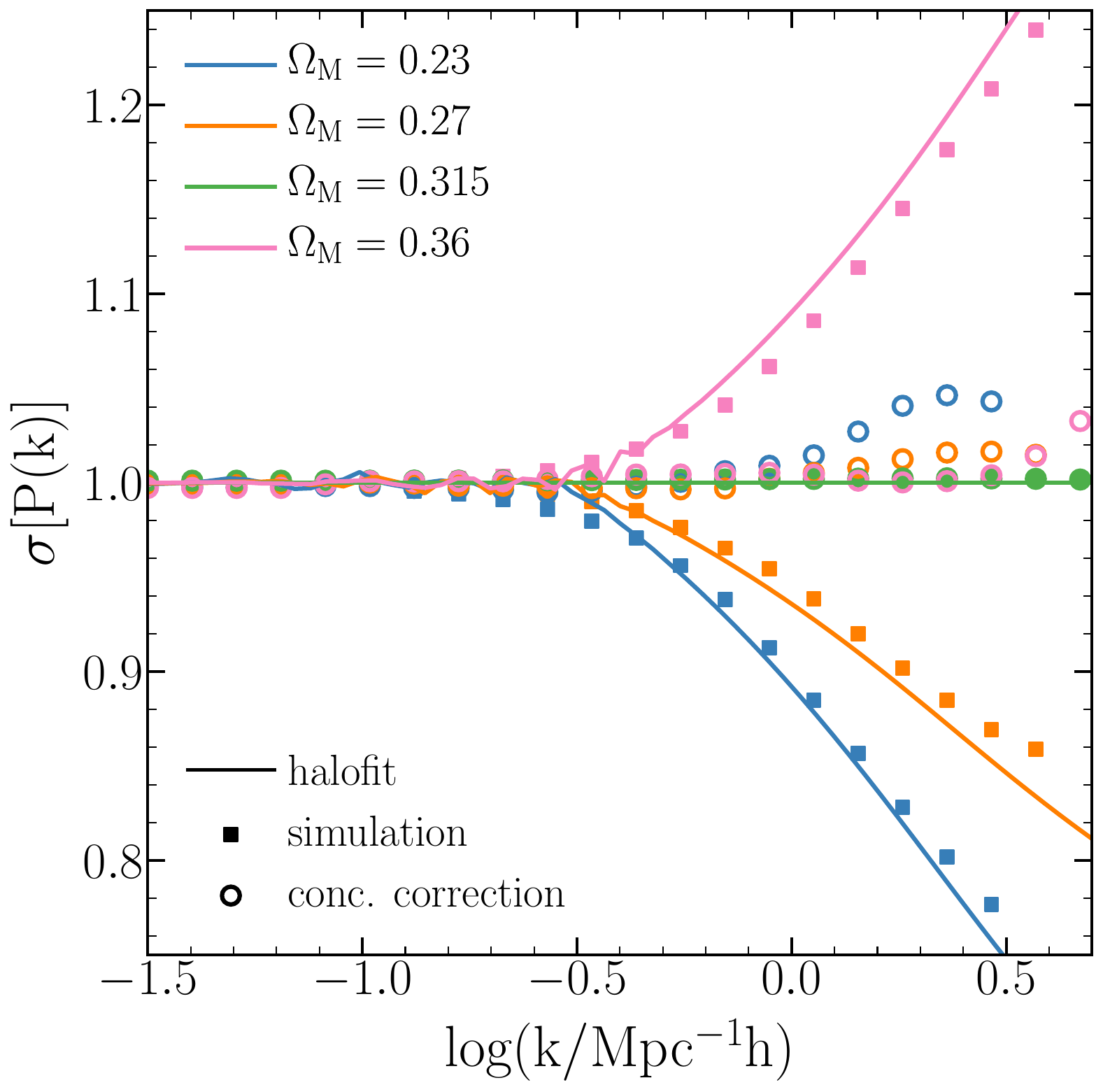}
\caption{The error associated to the cosmology scaling technique for the nonlinear mass $P(k)$ at $z=0$. We estimate this error as $\sigma[P(k)] \equiv P^{\rm scaled}(k)/P^{\rm direct}(k)$, where   ``scaled'' and ``direct'' superscripts denote quantities rescaled to, or computed directly in, the target cosmology, respectively. For each target cosmology, as indicated in the legend, we compute $\sigma[P(k)] $ either from $N$-body simulations (symbols) or using {\tt halofit} (solid lines). In all cases, the original cosmology is  $(\Omega_{\rm m}, \sigma_8) = (0.315, 0.9)$.}
\label{Fig:PredErr}
\end{figure}

The scaling algorithm is specifically designed so that quantities that depend on the amplitude of the linear fluctuation field are correctly reproduced. In the Press-Schechter formalism, the abundance and spatial properties of haloes fall within this category. In contrast, the algorithm makes no attempt to reproduce the growth history of fluctuations, thus quantities that depend on it, such as halo concentration and the power spectrum on very small scales, are expected to be predicted less accurately. 

Thus, the overall accuracy of the method will decrease the farther we scale away in cosmological parameter space from the original cosmology. We can see this in Fig.~\ref{Fig:PredErr} where we display the error of the scaling for the nonlinear mass power spectrum, P($k$), at $z=0$ for different cosmologies as scaled from $(\Omega_{\rm m}, \sigma_8) = (0.315, 0.9)$. In the top and bottom panels, different colours denote different target values of $\Omega_{\rm m}$ and $\sigma_8$, respectively, as indicated by the legend. For each case, we estimate the uncertainty in two ways: i) by rescaling the predictions of {\tt halofit} \citep[solid lines;][]{halofit} and ii) by comparing with simulations carried out using the target cosmology (symbols). We refer to \S\ref{sec:the_sims} for details on these simulations. 

Firstly, we can see that on scales larger than $\log (k/\ihMpc) \sim -0.5$, $P(k)$ is recovered almost perfectly. On smaller scales, rescaling can systematically over- or under-predict the non-linear clustering by up to 25\% in the case of $\Omega_{\rm m}$ or up to 10\% for $\sigma_8$. (Note that in \S\ref{sec:Improvements} we propose a method to dramatically reduce this error, which is shown as open symbols). Secondly, we can also see that the error amplitude is a monotonic function of the difference in cosmological parameters -- the larger the difference, the larger the error -- and that its magnitude is different for changes in different parameters. Nevertheless, the magnitude and shape of this systematic error are remarkably well predicted by {\tt halofit}. We will exploit these facts later in the paper to find the optimal point(s) to rescale from given a target region of cosmological parameter space.

\subsubsection{Computational Requirements}
\label{sec:comp_lim}

The main advantage of the scaling technique is that in a few seconds, and with the computational power of a standard laptop, one can scale a full simulation that would have taken days to run in a computer cluster. The only computationally expensive aspect is to carry out the original simulation, which, of course, needs to be done only once. 

\begin{figure}
\includegraphics[width=0.45\textwidth]{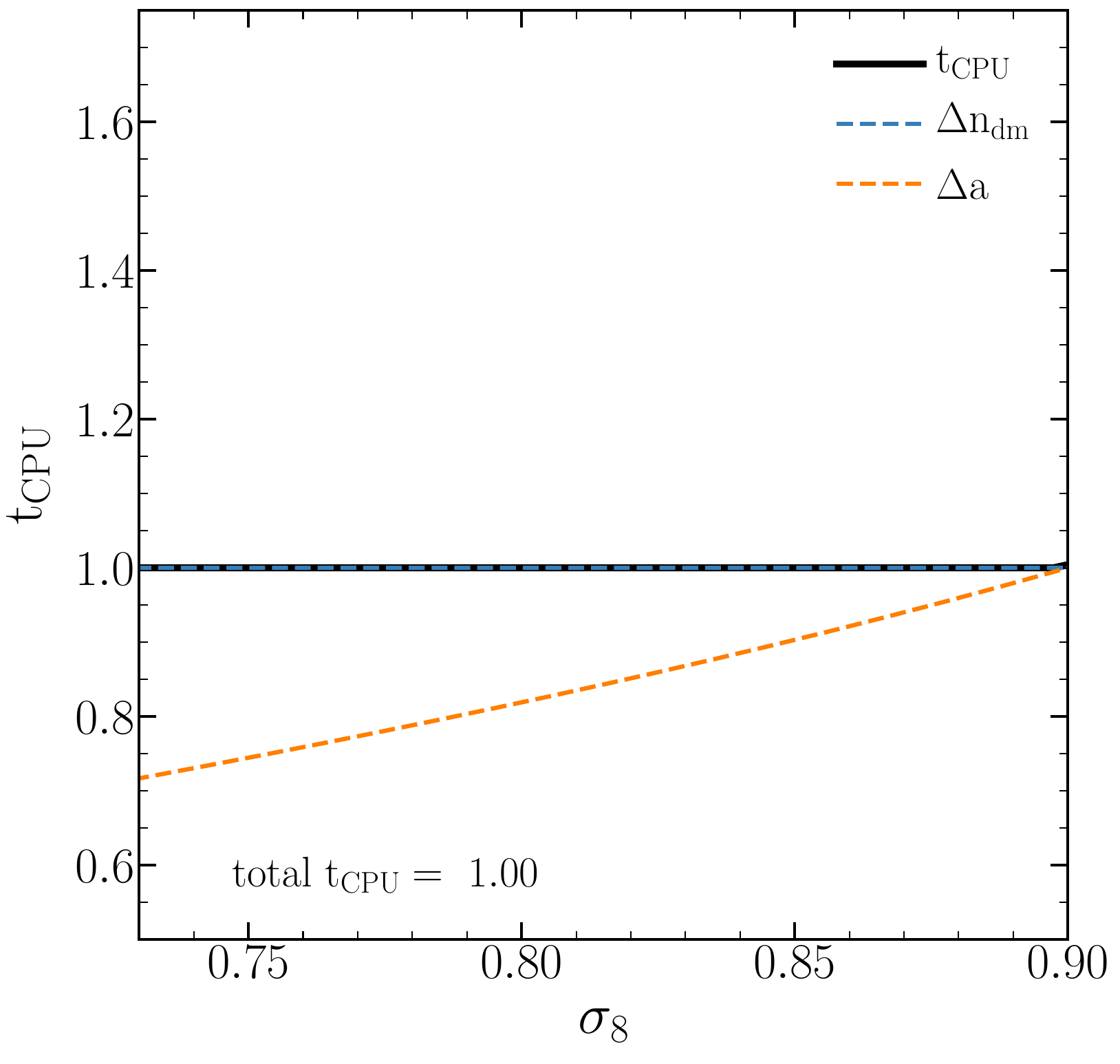}
\includegraphics[width=0.45\textwidth]{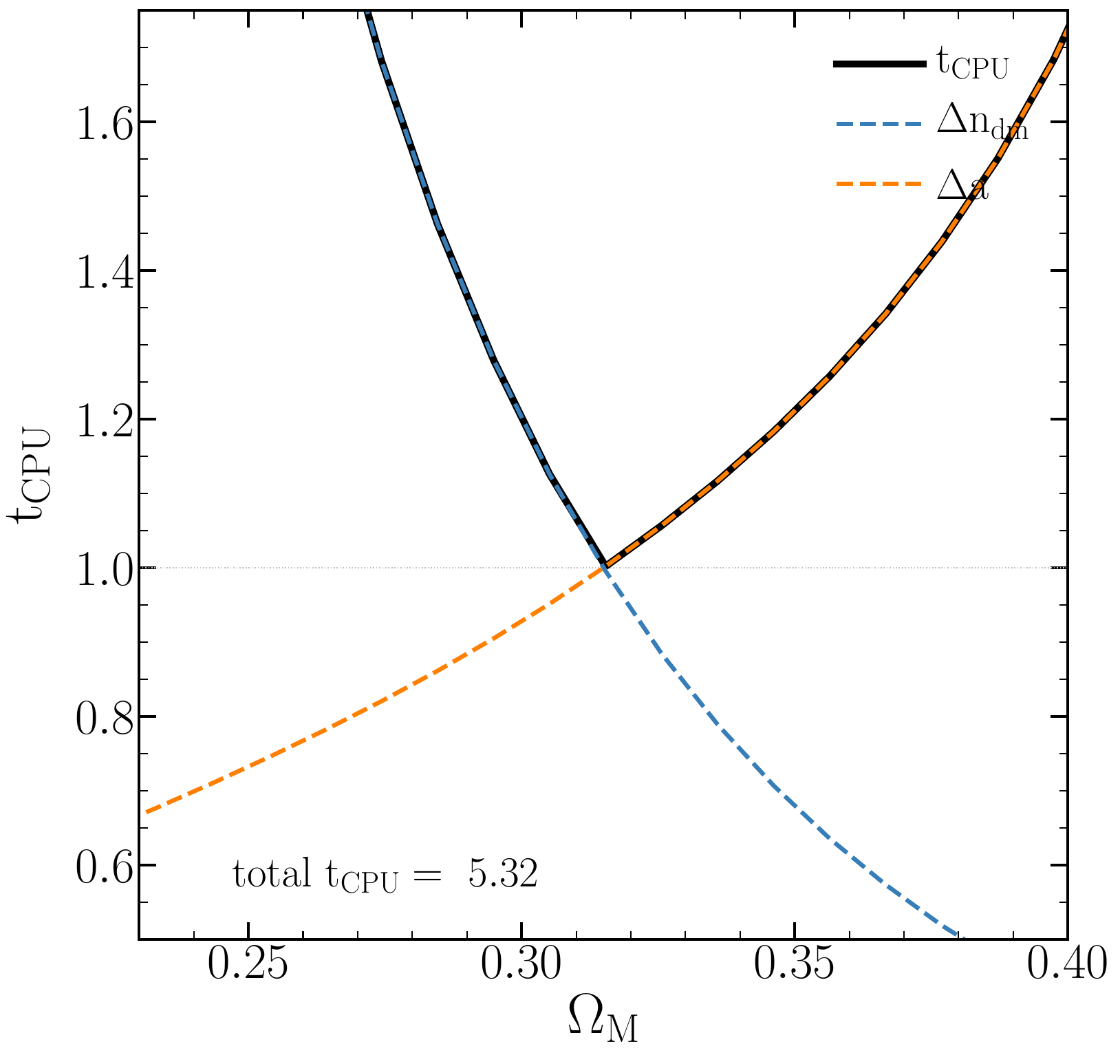}
\caption{The additional computational time necessary to run a simulation suitable for scaling as a function of the target cosmological parameter. The top panel shows changes in $\sigma_8$ and the bottom panel $\Omega_{\rm m}$. Orange dashed lines show the cost due to changes in the final expansion factor whereas blue dashed lines do so for the cost associated to the increase in the simulation volume (required for a given number density of objects). Black lines maximum computational cost for a given target cosmology.}
\label{Fig:Prop_CompTime}
\end{figure}

To obtain accurate rescaled predictions, however, the original simulation must fulfil several requirements (which will increase the computational time compared to a standard simulation):

\begin{itemize}

\item Simulating to the future: For some target cosmologies (typically with larger density fluctuations), the optimal expansion factor to scale from can be negative $\zstar < 0$. This requires the simulation to be evolved to the future, which increases its computational cost. 

As an example, orange lines in Fig.~\ref{Fig:Prop_CompTime} show the additional computational time required to scale to different values of $\sigma_8$ and $\Omega_{\rm m}$, from $(\Omega_{\rm m}, \sigma_8) = (0.315, 0.9)$. We estimate this CPU cost as proportional to the final expansion factor. For values lower than those in the original cosmology, $\zstar$ is located in the past, but for larger values, it is in the future, so the original simulation should be evolved longer, thus increasing the CPU time required.

\item Reaching the required number density: When the scaling length parameter, $s$, is larger than 1, it increases the simulation box length reducing the effective mass resolution. To achieve a given number density of haloes or subhaloes, the resolution of the original simulation should be higher so it compensates for the larger volume. 

Dashed blue lines of Fig~\ref{Fig:Prop_CompTime} show the additional computational time due to the increased volume (for a fixed number density of objects), assuming it scales as $s^3$. Since varying $\sigma_8$ does not change the fluctuation spectrum, $s=1$ and the simulation volume does not change. In contrast, small target $\Omega_{\rm m}$ values result in $s < 1$, thus the scaled volume is reduced and the computational time increases very rapidly. 

\item Large number of snapshots: In most cases, $\zstar$ will not coincide with any of the outputs of the original simulation. Thus, to maximize the accuracy of the method, the original simulation should produce a large number of snapshots. Although this is helpful for many applications, it increases the I/O and storage requirements in addition to the CPU time spent in group finders. % (e.g. {\tt SUBFIND} uses $\sim 15\%$ of the total computational time for $100$ snapshots).

%An alternative that we adopt here is to interpolate the results from two consecutive snapshots. With this we reduce de error coming from the finite number of snapshots to the sub percent level, when using $\sim 50$ outputs between $a=0.02$ and $a=1.25$. 

In summary, large changes in cosmology lead to a higher computational cost for the original simulation and also results in larger errors in the predictions. In the next section we will take advantage of these aspects to build an optimal suite of simulations to rescale from that maximizes accuracy and minimizes computational cost.

% FROM HERE
\end{itemize}

\section{Designing an Optimal suite of rescaling cosmologies}
\label{sec:Opt_Cosmo}

As mentioned in the previous section, as we scale to cosmologies farther away from the original, the computational cost of  the original simulation and the error on the scaling both increase. This implies that both limitations can be reduced simultaneously by designing a suite of multiple simulations to scale from. Each of these simulations would adopt different cosmologies and together they would cap the computational cost and the maximum error of the scaling method. We explore this idea next.

\begin{figure*}
\includegraphics[width=\textwidth]{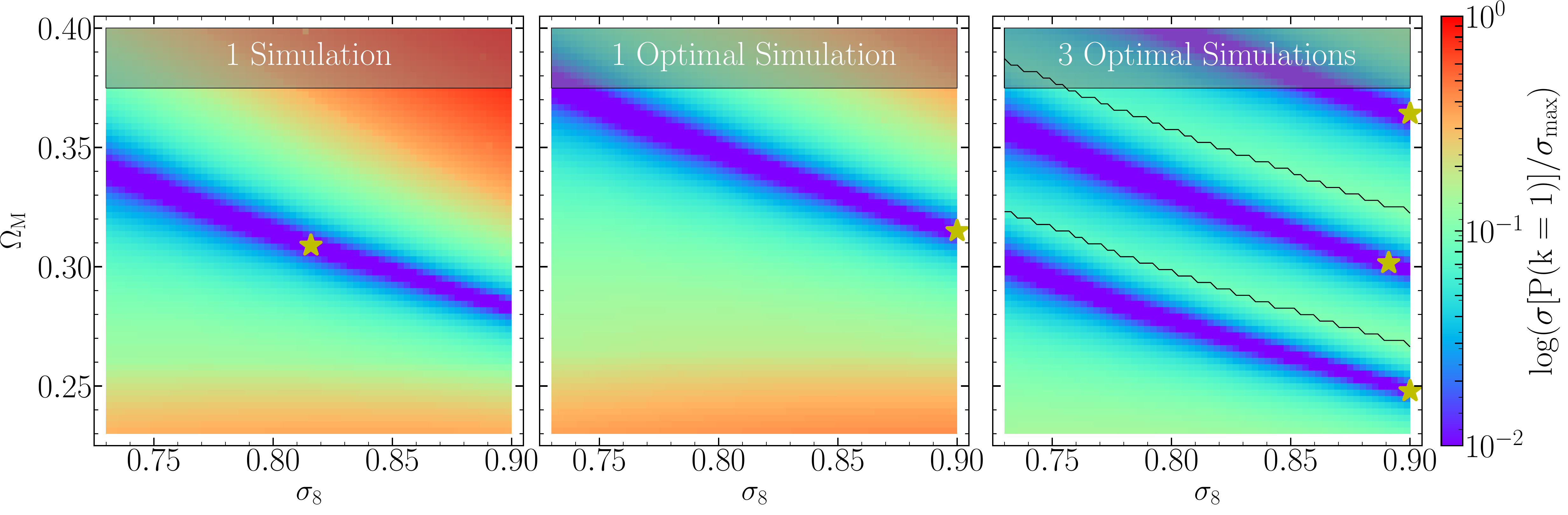}
\caption{(left) The error associated to the cosmology scaling ($\sigma[P(k)]$) at $k=1 \ihMpc$ by scaling a cosmology with $\sigma_8=0.8159$ and $\Omega_{\rm m}=0.3089$ (\citealt{Planck13_16}, showed as a yellow star) to the full range of cosmologies. The values are normalized by the maximum scaling error. (middle) Same as the left panel but for an original cosmology of $\sigma_8=0.9$ and $\Omega_{\rm m}=0.315$, the most computationally efficient single cosmology. (right) Same as the left and middle plot, but using three cosmologies to scale from instead of one ($\sigma_8$, $\Omega_{\rm m}$ = (0.9, 0.248), (0.89, 0.301) and (0.9, 0.364)). The solid line represents the areas where each of the main cosmologies is used to scale from.}
\label{Fig:NPerf}
\end{figure*}

First, we define the range of cosmological parameters of interest. We consider 8 cosmological parameters:

\begin{eqnarray}
\label{eq:par_range}
\sigma_8                  &\in& [0.73, 0.9]\\
\Omega_{\rm m}            &\in& [0.23, 0.4]\\
\Omega_b                  &\in& [0.04, 0.06]\\
n_s                       &\in& [0.92, 1.01]\\
h\,[100\,{\rm km}\,{\rm s^{-1}} {\rm Mpc^{-1}}]  &\in& [0.6, 0.8]\\
M_{\nu}\,[{\rm eV}]        &\in& [0.0, 0.4]\\
w_{0}                     &\in& [-1.15, -0.85]\\
w_{a}                     &\in& [-0.3, 0.3]
\end{eqnarray}

\noindent where $M_{\nu}$ is the total mass in neutrinos, and $w_0$ and $w_a$ define the time evolution of the dark energy equation of state: $w(z) = w0 + (1-a) w_a$. We set the range for $(\sigma_8, \Omega_{\rm m}, \Omega_b, n_s)$, as a $\sim 10\ \sigma$ region around the best fit parameters of the analysis of \cite{Planck13_16}. For the Hubble parameter, $h$, we expand the range to cover a $\sim 4 \sigma$ region around current low-redshift measurements from supernovae data \citep{Riess:2019}. 

For a given original cosmology, $\bm{\theta_{o}}$, we employ a proxy for the accuracy of scaling from it, $\varepsilon[\bm{\theta_o}]$. We define this quantity as the average rescaling uncertainty for the nonlinear power spectrum at $z=0$ over our whole parameter space:

\begin{equation}
\varepsilon[\bm{\theta_{o}}] \equiv \langle \sigma[P(k=1)](\bm{\theta_i}) \rangle
\label{ref:error}
\end{equation} 

Operationally, we compute $\sigma[P(k=1)](\bm{\theta_i})$ employing {\tt halofit} at $z=0$ and perform the average by evaluating it over 3125 random target cosmologies. We then estimate the computational cost of a simulation carried out adopting $\bm{\theta_{o}}$ as:

\begin{equation}
t_{\rm CPU}[\bm{\theta_{o}}] \propto a_{\rm max} \times s^3_{\rm max}
\end{equation} 

\noindent where $a_{\rm max}$ and $s_{\rm max}$ are the maximum original expansion factor and scaling length parameters evaluated over the same $3125$ trial cosmologies.

We now consider the case where we the employ a {\it set} of $N$ original cosmologies to scale from. In this case, we estimate the combined CPU cost and rescaling uncertainty, $\varepsilon^{(N)}[\{\bm{\theta}\}]$ and $t^{(N)}_{\rm CPU}[\{\bm{\theta}\}]$ as follows. $t_{\rm CPU}^{(N)}$ is simply given as the sum of the individual CPU cost of the simulations in the set, whereas $\varepsilon^{N}$ is given by the average of the minimum uncertainty of each simulation separately at every trial cosmology.

Before searching for an optimal combination of simulations, we illustrate the previous points in Fig.~\ref{Fig:NPerf}. In the left panel we consider a cosmology (indicated by the yellow star) and display the uncertainty $\sigma[P]$ of rescaling it as a function of target ($\sigma_8$, $\Omega_{\rm m}$) values. We can clearly see that there are regions over which the scaling has similar performance (roughly along $\sigma_8 \,\Omega_{\rm m}$ degeneracy line). In the middle panel we show the same quantity but for the cosmology that minimizes the average uncertainty (eq. 9). 

This cosmology prefers roughly the same value for $\sigma_8$ as before, but a much larger value for $\Omega_{\rm m}$. In this way, the extreme values of uncertainty are reduced while avoiding simulating very large boxes.
In the right panel, we display the expected accuracy from a set of three simulations that minimize the global average uncertainty, $\sigma^N$. 
We can see that in this case the maximum uncertainty is capped and we achieve good precision everywhere in cosmological space. 

We point out that the precision of the scaling is not necessarily better when only one cosmological parameter is modified. Instead, the best performance is achieved along some particular degeneracies of the parameters, as can be seen in Fig.~\ref{Fig:NPerf}.

\begin{figure}
\includegraphics[width=0.45\textwidth]{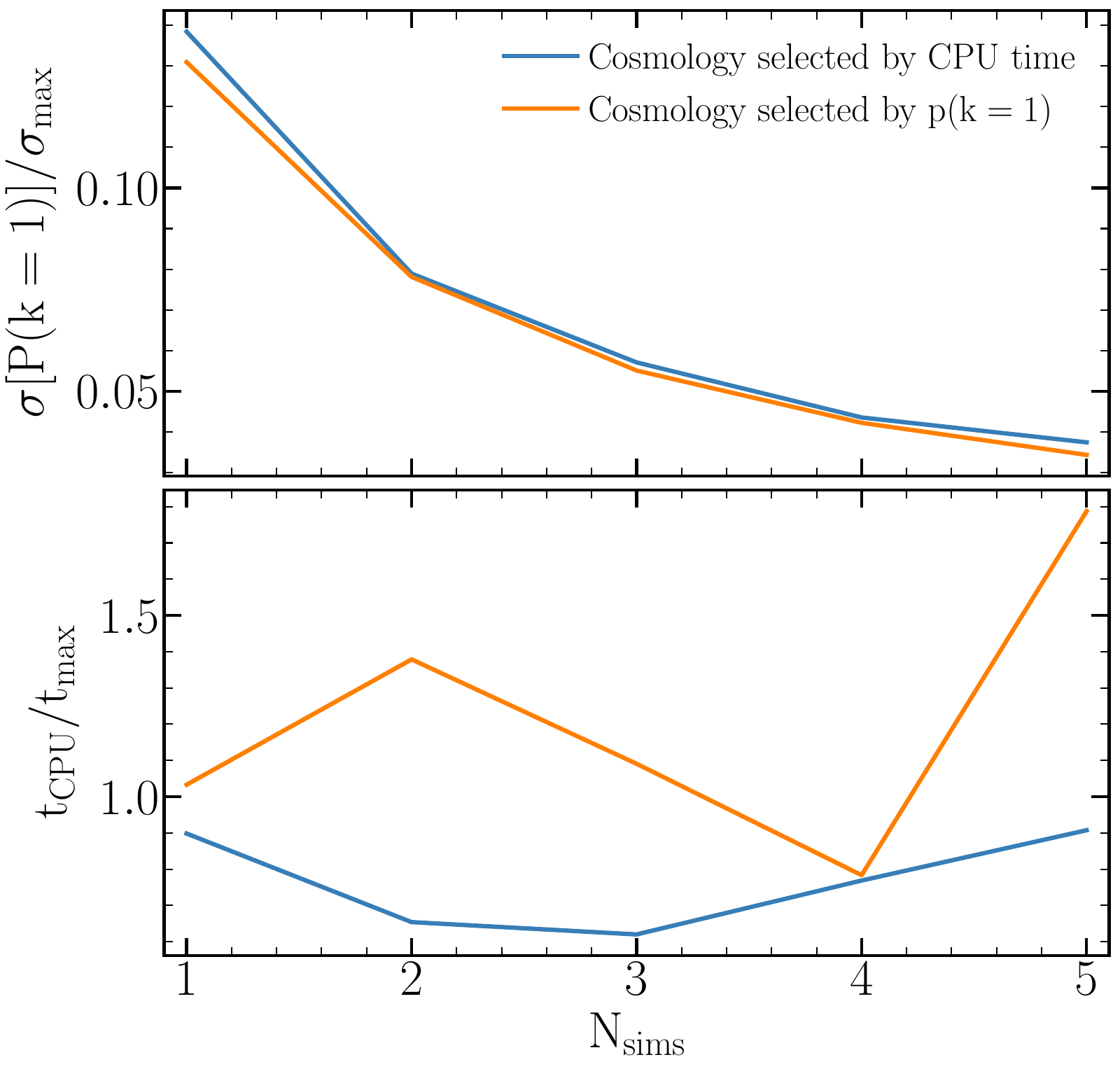}
\caption{Computational performance and predicted error of the dark matter power spectrum at $k {\rm =1\ \ihMpc}$ as a function of the number of simulation used for scaling our cosmological parameter space. The optimal number of simulations in terms of computational time and performance is between 2 and 3 simulations. The values are normalized by the scaling time and error of a single simulation with a Planck-13 cosmology.}
\label{Fig:BestN}
\end{figure}

To define an optimal suite of simulations to rescale from, we first created $\sim 10^9$ sets of $1$ to $5$ simulations with different combinations of $\sigma_8$, $\Omega_{\rm m}$, $\Omega_{\rm b}$, $n_s$, and $h$ (and fixing $M_{\nu}=0$, $w_0=-1$, and $w_a=0$) and evaluate $\varepsilon[\bm{\theta_{o}}]$. We then identified the sets that either minimize the total computational time or the global scaling accuracy. In Fig.~\ref{Fig:BestN} we show the computational time and the mean error of these sets, as a function of the number of simulations in it.

Firstly, we see that the overall error decreases as we consider more 
simulations in the set, with the sets selected according to $\varepsilon$ or $t_{\rm CPU}$ delivering almost identical performances. Interestingly, this increase in precision does not necessarily imply a larger amount of CPU time. In fact the set selected using $t_{\rm CPU}$ requires resources almost independent of $N_{\rm sims}$, with a minimum at $N_{\rm sims} \sim 2-3$. Although the precision can be increased further by considering higher $N_{\rm sims}$, the I/O and storage requirements increase considerably. 

Therefore, we identify a set of 3 simulations selected according to $t_{\rm CPU}$ as the optimal suite for the cosmology rescaling method. The cosmological parameters of this set are provided in Table~\ref{Table:CosmoParams}. We name these cosmologies \textsc{Vilya}, \textsc{Nenya} and \textsc{Narya}.\footnote{\textsc{Vilya}, \textsc{Nenya} and \textsc{Narya} are the three ``Rings of Power" given to the Elves of Middle-earth in the ``Lord of the Rings" mythology.} Notice that these cosmologies are homogeneously distributed in $\Omega_{\rm m}$ (that is the cosmological parameter that produces the largest uncertainties on the scaling and the highest additional computational cost). The other properties prefer more extreme values. This is to compensate for changes on the length and time scaling parameters, minimizing the additional computational time without compromising accuracy. We would like to point out that these cosmologies will not only be used in the framework of this work, but also in a set of very large simulations designed to make accurate predictions for nonlinear clustering as a function of cosmology \citep{Angulo:2020}.

\begin{table}
\caption{
The cosmological parameters of the three main cosmologies used in this paper: Vilya, Nenya and Narya, and the parameters of the best fit of the first data release of Planck \citep{Planck13_16} }
\begin{center}
 \begin{tabular}{c c c c c c c c c}
 \hline
 & $\rm \sigma_8$ & $\rm \Omega_{\rm m}$ & $\rm \Omega_b$ & $\rm n_s$ & $\rm h$ & $M_{\nu}$ & $w_0$ & $w_a$\\ 
 \hline
 Vilya & 0.9 & 0.27  & 0.06 &  0.92 & 0.65 & 0.0 & -1.0 & 0.0   \\ 
 Nenya & 0.9 &  0.315 & 0.05 &  1.01 & 0.60 & 0.0 & -1.0 & 0.0   \\ 
 Narya & 0.9 & 0.36  & 0.05 &  1.01 &  0.70 & 0.0 & -1.0 & 0.0   \\ 
 Planck-13 & 0.8288 & 0.3071  & 0.04825 &  0.9611 & 0.6777 & 0.0 & -1.0 & 0.0   \\   
\end{tabular}
\end{center}
\label{Table:CosmoParams}
\end{table}

\section{The performance of the scaling}
\label{sec:main}

In previous sections, we showed that a suite of three cosmologies represent an optimal choice  in terms of computational resources and accuracy. In this section, we carry out a suite of $N$-body simulations adopting those exact three cosmologies and then we empirically measure the accuracy of the scaling technique. 

In \S\ref{sec:the_sims} we provide details on the $N$-body code and our sets of original and target simulations. In \S\ref{sec:Improvements} we discuss how to further improve the scaling. In \S\ref{sec:accuracy} we present the performance of the scaling in terms of mass, halo, and subhalo power spectra.

\subsection{The simulations}
\label{sec:the_sims}
\subsubsection{$N$-body code}
\label{sec:nbody_code}

All of our $N$-body simulations were carried out using an updated version of {\tt L-Gadget3} \citep{Angulo:2012} -- a lean version of {\tt GADGET} \citep{Springel:2005} used to run the Millennium XXL simulation. This version of the code creates the initial conditions on-the-fly using 2nd order Lagrangian Perturbation Theory, and allows an on-the-fly identification of haloes and subhaloes using a Friend-of-Friend algorithm \citep[{\tt FOF}][]{Davis:1985} and an extended version of {\tt SUBFIND} \citep{Springel:2001}.

Our updated version of {\tt SUBFIND} can better identify substructures by considering information of its past history, while also measuring properties that are non-local in time such as the peak halo mass ($\rm M_{peak}$), peak maximum circular velocity ($\rm V_{peak}$), infall subhalo mass ($M_{\rm infall}$), and mass accretion rate among others. We refer to Angulo et al. in prep. for further details.

An additional feature we include in {\tt L-Gadget3} is the ability to store a diluted sample of dark matter particles. This sample is constructed by uniformly selecting 1 every $4^3$ particles in Lagrangian space. The power spectrum of the full particle distribution and this diluted sample agree at the sub percent level up to $k\sim5\,\ihMpc$, while substantially reducing I/O and storage requirements of the simulations. Throughout this paper, we will employ this catalogue when exploring dark matter clustering. 

Power spectra were computed using Fast Fourier Transforms on a density mesh using Triangular Shape Cloud mass assignment scheme with $512^3$ points. The Fourier meshes were then interlaced to reduce the noise associated with aliasing and the finite mesh \citep{Sefusatti:2016}. This ensures we obtain sub percent accurate power spectra estimations up to $k \sim 6\,\ihMpc$. Note we do not subtract the Poisson shot noise from our measurements.

\subsubsection{The target cosmologies}
\label{sec:target_cosmo}

We ran two sets of simulations. The first one corresponds to the optimal rescaling suite, which consists of 3 simulations --  \textsc{Vilya}, \textsc{Nenya}, and \textsc{Narya} -- defined in the previous section (c.f. Table 1). The second set consists of a suite of simulations with different cosmologies with which we will test the performance of the rescaling suite. Specifically, we sample uniformly the range given by Eq.~\ref{eq:par_range} with 5 simulations per parameter, while keeping the others fixed to those of \textsc{Nenya} (our central simulation).

The suite of original simulations followed $768^3$ particles in a $\rm (256\,\hMpc)^3$ cubic periodic volume. We have stored 100 snapshots between $z=49$ and $z = -0.2$ (i.e. a = 1.25, to use when $a^{*} > 1$ ). The resolution was chosen to ensure completeness in a subhalo sample with a number density of $\rm 0.01\ h^3Mpc^{-3}$ selected by $V_{\rm peak}$ even when scaling to extreme cosmologies with a large length rescaling parameter ``$s$'' (i.e. lower maximum number density). 

The suite of target simulations has the same number of particles but within slightly different volumes, so that their box size matches that of the original simulations after rescaling to the respective cosmology. We have stored 70 snapshots distributed between $z=49$ and $z = 0$. 

To reduce cosmic variance and allow a more accurate comparison, the initial conditions of all simulations have identical white noise fields and have been ``paired-and-fixed'' following \cite{Angulo:2016}. 
In total, we have carried out $2\times35$ test simulations, with additional runs to quantify resolution and other numerical checks.

Complementing these simulations, we have carried out another simulation adopting a cosmology compatible with that preferred by \cite{Planck13_16} (cosmological parameters are provided in Table~1). For this particular simulation, we did not match the simulation side-length, nor fixed its initial Fourier amplitudes. We do this since we will employ this simulation to build a mock catalogue to study the ability of our approach to constrain cosmological parameters. Thus, we choose to keep the intra-simulation variance compatible with real observations.

\subsection{Further improvements to the scaling}
\label{sec:Improvements}

\begin{figure}
\includegraphics[width=0.45\textwidth]{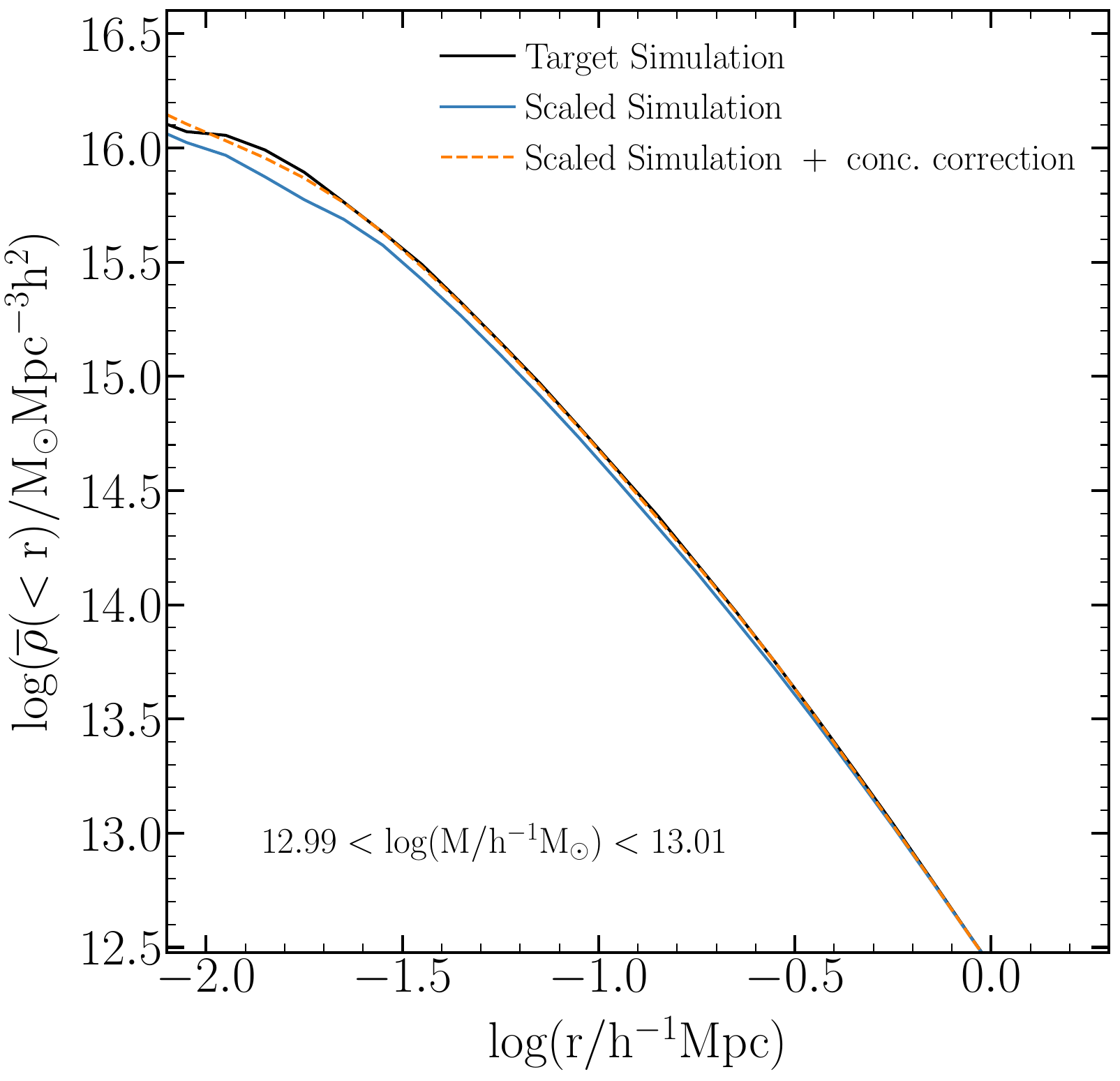} 
\includegraphics[width=0.45\textwidth]{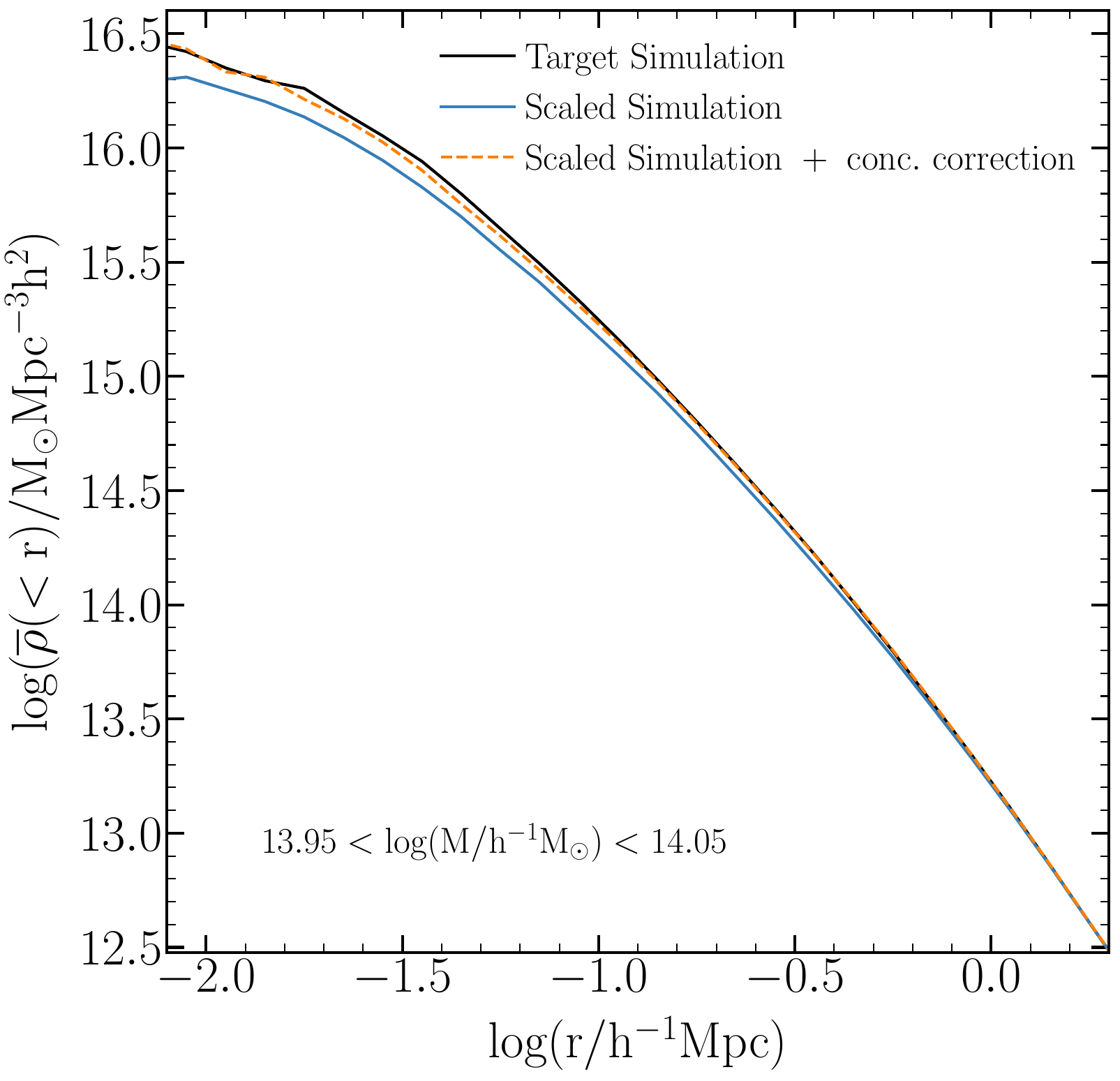} 
\caption{Density profile of haloes with mass $\sim 10^{13}h^{-1} {\rm M_{\odot}}$ (top) and $\sim 10^{14}h^{-1}{\rm M_{\odot}}$ (bottom). Black lines show the measurements for haloes in a $N$-body simulation with a Nenya-like cosmology with $\sigma_8 = 0.73$, whereas solid blue and dashed orange lines show the result for haloes in a scaled simulation before and after the differences in the expected concentration-mass-redshift relation is accounted for. See \S\ref{sec:Improvements} for more details.}
\label{Fig:Profile}
\end{figure}

Whereas the accuracy on the rescaled power spectrum is better than $1\%$ on large scales (c.f. Fig.~\ref{Fig:PredErr}), it degrades rapidly on smaller scales, reaching up to $20\%$. This is simply a consequence of a mismatch in the concentration mass relation owing to different formation times in the rescaled and target cosmologies. 

To improve the accuracy of the scaling, we developed a correction that takes advantage of recent models for the cosmology dependence of the concentration-mass relation. This correction consists on slightly perturbing the distance of particles to the centre of their host halo ($r$) as follows:

\begin{equation}
  r \rightarrow r + \Psi(r | \{ M, c(M)\}),
\end{equation}

\noindent where $M$ is the host halo mass, and $\Psi(r)$ is a displacement field defined implicitly from the difference in the cumulative mas profiles:

\begin{equation}
\begin{split}
    \int^{r + \Psi(r)}{\rm d}^3y\,\rho(y,M',c'(M))= \int^{r}{\rm d}^3y\,\rho(y,M,c(M))
\end{split}
\end{equation}

\noindent where (un)primed quantities refer to those evaluated in the (target) scaled cosmology. Operationally, we assume $\rho$ to be given by an NFW form \cite{NFW:1995}, and the concentration mass relation, $c$, to be that predicted by \citealt{Ludlow:2016}. 
Note we do not force the individual profile of each halo to follow an NFW profile, but instead assume that the cosmology-dependence of the profile does. 

\begin{figure*}
\includegraphics[width=\textwidth]{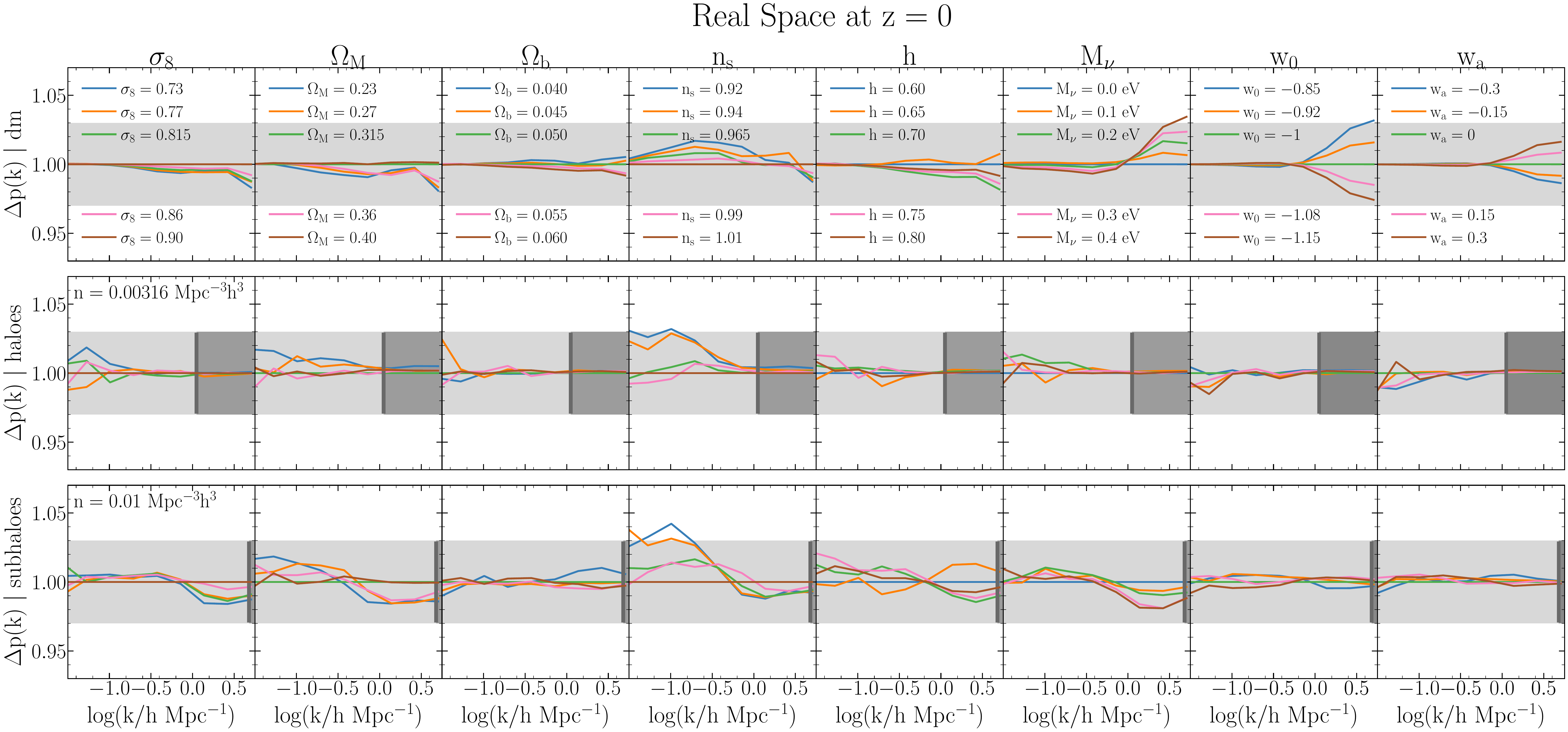}\\ 
\vspace{1cm}
\includegraphics[width=\textwidth]{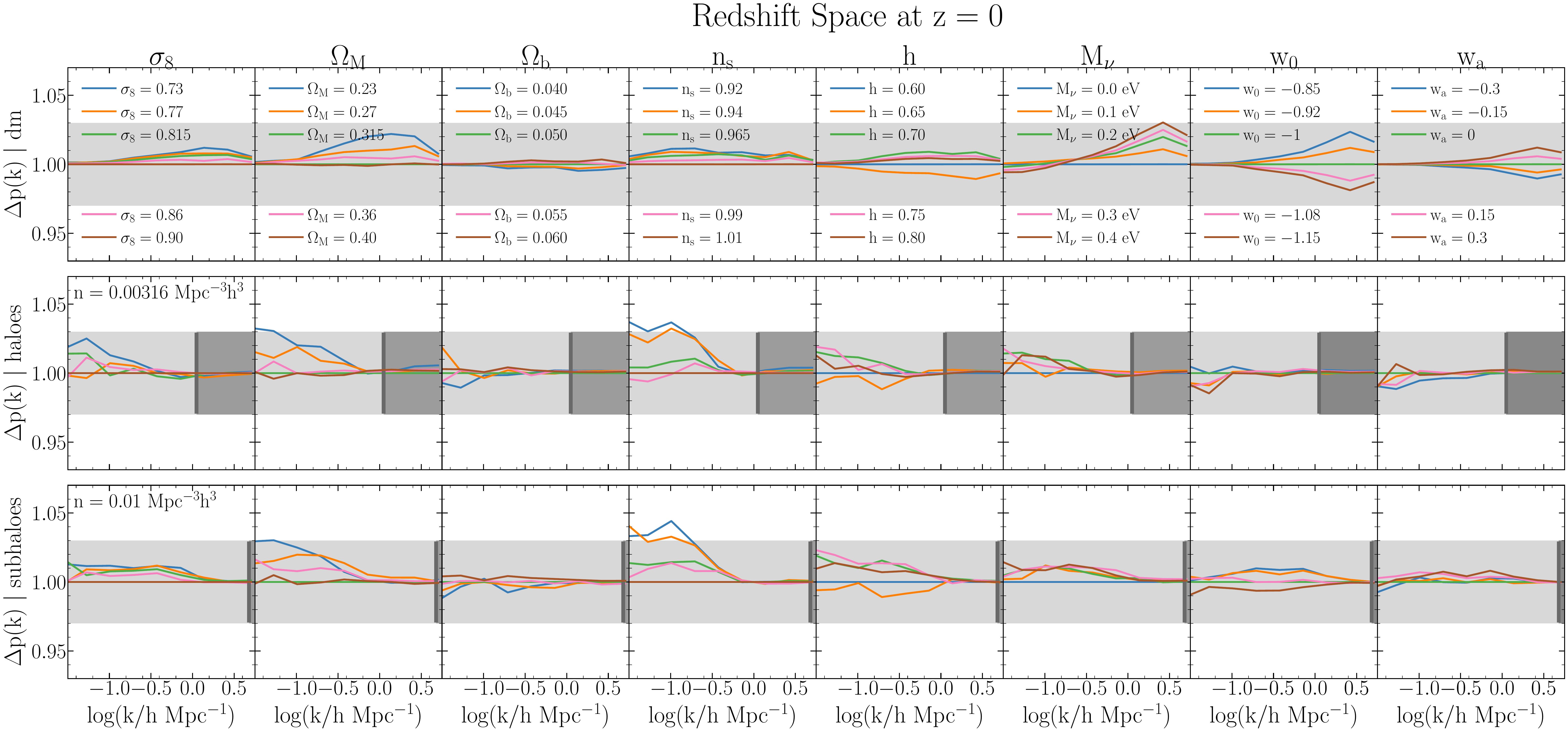}
\caption{The ratio between the power spectra of scaled and target simulations in real and redshift space (top and bottom panels respectively). The target cosmologies adopt Nenya cosmologies varying one cosmological parameter at a time, as indicated in the label of each figure. The eight columns, from left to right, show changes in $\sigma_8$, $\Omega_{\rm m}$, $\Omega_b$, $n_s$, $h$, $M_{\nu}$, $w_0$, and $w_a$. The light shaded region highlights a $3\%$ discrepancy. The top row displays results for dark matter, whereas middle and bottom rows do so for haloes with a number density of $\rm 3.16\times10^{-3}\,h^3Mpc^{-3}$ selected according to their mass, and subhaloes with a number density of $\rm 10^{-2}\ h^3Mpc^{-3}$ selected according to their peak maximum circular velocity ($\rm V_{peak}$). The dark shaded region shows the region where Poisson shot-noise is above than 80\% of the power spectrum amplitude.}
\label{fig:full_scaling}
\end{figure*}

Fig.~\ref{Fig:Profile} shows this ``concentration correction'' in practice. We display the density profile of haloes with $M \sim 10^{13}\,h^{-1}{\rm M_{\odot}}$ (top panel) and with  $M \sim 10^{14}\,h^{-1}{\rm M_{\odot}}$ (bottom panel). Black lines show the profile as measured in one of our target simulations with $\sigma_8=0.73$, whereas blue indicates the profile in our rescaled simulation. We take advantage of the fact that our simulations share phases in their initial conditions to identify the same haloes in both simulations. We do so by crossmatching the position and masses of haloes in both simulations, which results in $377$ matched haloes for the low halo mass sample and $145$ for the high mass one. 

We can see that rescaled haloes are systematically less concentrated (thus we expect a lower amplitude in the power spectrum, consistent with the blue lines in the top panel of Fig.~\ref{Fig:PredErr}). The profiles after concentration correction, displayed by dashed lines, capture this effect and correct the profiles remarkably well, on all scales down to 10 - 15 $\hkpc$ (a few times the softening length of our simulations). Although not shown here, other halo masses and concentration-mass relations show similar performances.\footnote{Considering all changes in cosmology, \citealt{Ludlow:2016} is the model yielding the best overall performance but \cite{diemer:2018} is the most accurate at higher masses.} 

Open symbols in Fig.~\ref{Fig:PredErr} show the power spectrum after this correction. We can see that now the uncertainty in the scaling is below $\sim3\%$ over all wavelengths, even for extreme changes in cosmologies. In the next subsection we will explore systematically the accuracy as a function of changes in different cosmological parameters.

\subsection{Accuracy on the clustering predictions}
\label{sec:accuracy}

We now test the performance of the scaling technique applied to our optimal simulation suite. We consider a wide range in target cosmologies, covering $\sigma_8$, $\Omega_{\rm m}$, $\Omega_b$, $n_s$, $h$, $M_{\nu}$, $w_0$ \& $w_a$, for which we will compare direct $N$-body simulations and rescaling predictions. We refer to \S\ref{sec:the_sims} for details on the simulations we employ.

For every target cosmology, we identify and employ the simulation in the original suite that would return the smallest error on the non-linear power spectrum at $k=1\,\ihMpc$. Since this could be a CPU intensive process, we trained a classification Artificial Neural Network (a.k.a ANN) using {\tt
scikit-learn}\footnote{\url{https://scikit-learn.org}}, an open-source tool for data mining and data analysis for python. For the training sample, we feed the ANN with 10,000 target cosmologies and the optimal original cosmology to scale from using the theoretical prediction of the error of the scaling given by {\tt halofit} (see \S\ref{sec:halomodel} for more details). The misclassification error in the trained ANN prediction is less than $2\%$ (normally for cosmologies that are in the limit of two main cosmologies to be scaled from) and its execution time is less than 4 milliseconds for a single CPU. In practice, for all the target cosmologies except the most extreme ones ($\Omega_{\rm m} = 0.4,\ \Omega_{\rm m} = 0.36,\ h = 0.8,\ h = 0.75\ \&\ h = 0.7$), \textsc{Nenya} will be the original simulation of choice. This also means that, except from these 5 target cosmologies, we only test the change of a single parameter at a time. As it was mentioned in section~\ref{sec:Opt_Cosmo}, changing a single parameter does not necessarily achieves a better precision compared to changing many properties at the same time. We chose this selection of target cosmologies to improve the understanding of scaling each particular cosmological parameter. In Appendix A, we test the performance of the scaling when choosing target cosmologies over the whole parameter space, finding similar performances as those shown here.
% , and while it somehow fail to cover the most extreem corners of our parameter space, it still cover a large region of it. Because of this, we expect that the presition of the scaling in these cosmologies to be a good representative of the presition in all the cosmological parameter space.}

In addition to testing the performance for the dark matter power spectrum, we also test the power spectrum of haloes and subhaloes. We select the haloes (subhaloes) with a fix number density of $n=0.00316\ (0.01)\ h^{3} {\rm Mpc}^{-3}$ based on their mass (peak maximum circular velocity). The particular number density is the densest limit that ensures objects are properly resolved given our resolution while minimizing shot noise. Nevertheless, and although not shown here, we have verified our results are representative of those at other thresholds. Note that for the scaling of haloes and subhaloes, we use a mass-dependent displacement (i.e. applying the displacement field of the matter component, weighted by the mass dependent bias of each structure) following the procedure of \cite{Mead:2014}.
 
The ratio between the power spectra measured in the target and scaled simulations for dark matter particles, haloes and subhaloes at $z=0$ are shown in Fig.~\ref{fig:full_scaling}. We chose 4 different and extreme cosmologies per cosmological parameter to scale to, plus scaling the simulation to itself as a sanity check. The ratios are shown until the smallest scale our grid allows. The region where the expected shot-noise reaches $>80\%$ of the signal is indicated by the dark-grey shaded region. The top figure displays results in real space and the bottom in redshift space.

Overall, we find that the scaling provides remarkably accurate predictions. For the mass power spectrum, it is accurate to better than $\sim3\%$ from very large to highly non-linear ones. To put this in context, we recall that the accuracy among different $N$-body codes is at $\sim 3\%$ at $\log (k/\ihMpc) \sim -0.5$ (i.e. \citealt{Schneider:2016}, note this is mostly driven by the accuracy in the time integration and force calculation). Note that the accuracy is somewhat poorer for non-standard $\Lambda$CDM parameters ($M_{\nu}$, $w_0$, and $w_a$). We have tracked this to inaccuracies in the models for concentration-mass relations, as they are typically less tested and calibrated for changes in those cosmological parameters.

In the case of $V_{\rm peak}$-selected subhaloes, the accuracy remains very high -- better than $4 \%$ for all target cosmologies and scales considered. To put again this value into context, we note that galaxy assembly bias can cause differences of up to 20\% in the 2-halo term of the galaxy correlation function, with non-trivial dependences on redshift and the type of galaxy selection \citep{C19}. Thus, the rescaling uncertainty is $\sim4-5$ times smaller than the impact of assembly bias, which also incidentally implies the rescaling is correctly capturing most of it. Since galaxy assembly bias is a direct consequence of halo assembly bias, these results suggest our method is recovering most of the halo assembly bias signal of the simulation.

There are, however, hints of systematic differences -- most notably for large changes in $n_s$, which is the cosmology that presents the largest changes in the original and target primordial power spectra.
There are various plausible explanations for this systematic error. Firstly, as we discussed earlier, the formation histories differ between the original and target haloes. This implies that subhaloes were accreted at different times and thus the probability of them to be tidally disrupted differs in the original and target cosmologies. This, in principle, could be accounted for by tracking the descendants of subhaloes tidally disrupted (a.k.a. orphan subhaloes). Another possibility is the response of halo formation on the background overdensity: changes in the large scale power makes halo formation more or less efficient in different regions. This effect is currently not modelled in the rescaling algorithm and should modify the large-scale clustering of objects. We defer to the future the investigation of these two issues, which can potentially enhance the accuracy of the method even further. 

\begin{figure*}
\includegraphics[width=\textwidth]{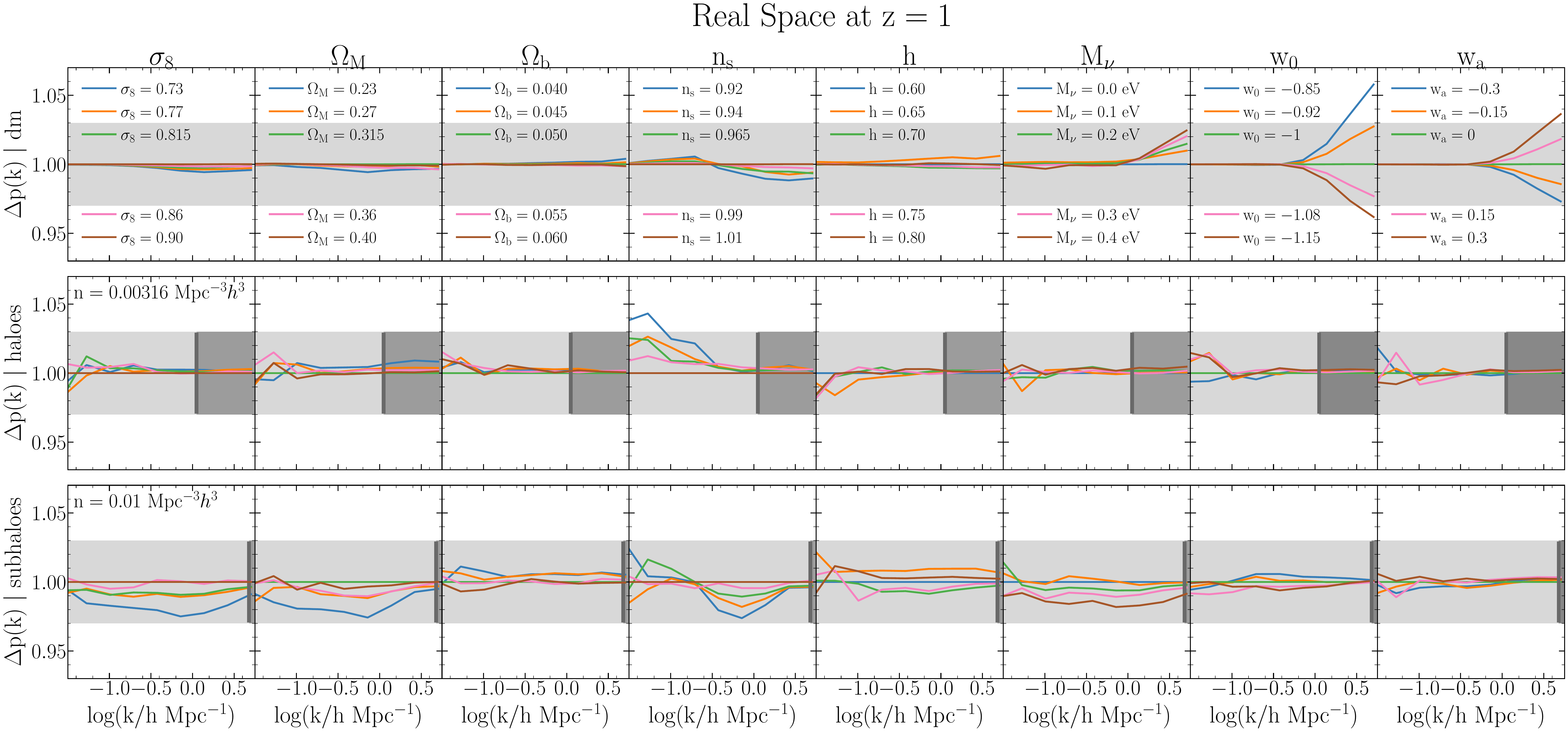}\\
\vspace{1cm}
\includegraphics[width=\textwidth]{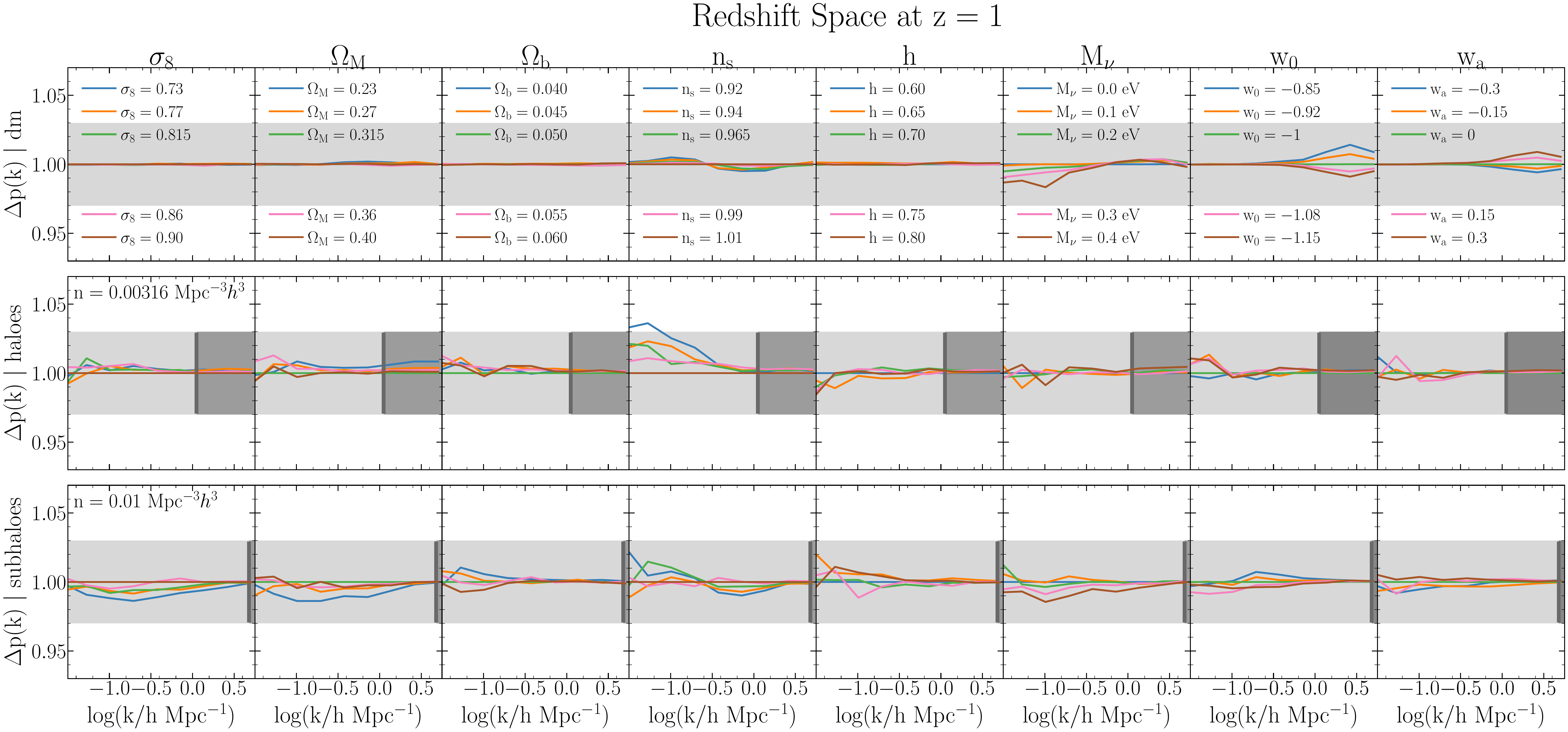}
\caption{Same as fig.~\ref{fig:full_scaling} but for $\rm z=1$}
\label{fig:full_scaling_z1}
\end{figure*}

\begin{figure*}
\includegraphics[width=0.8\textwidth]{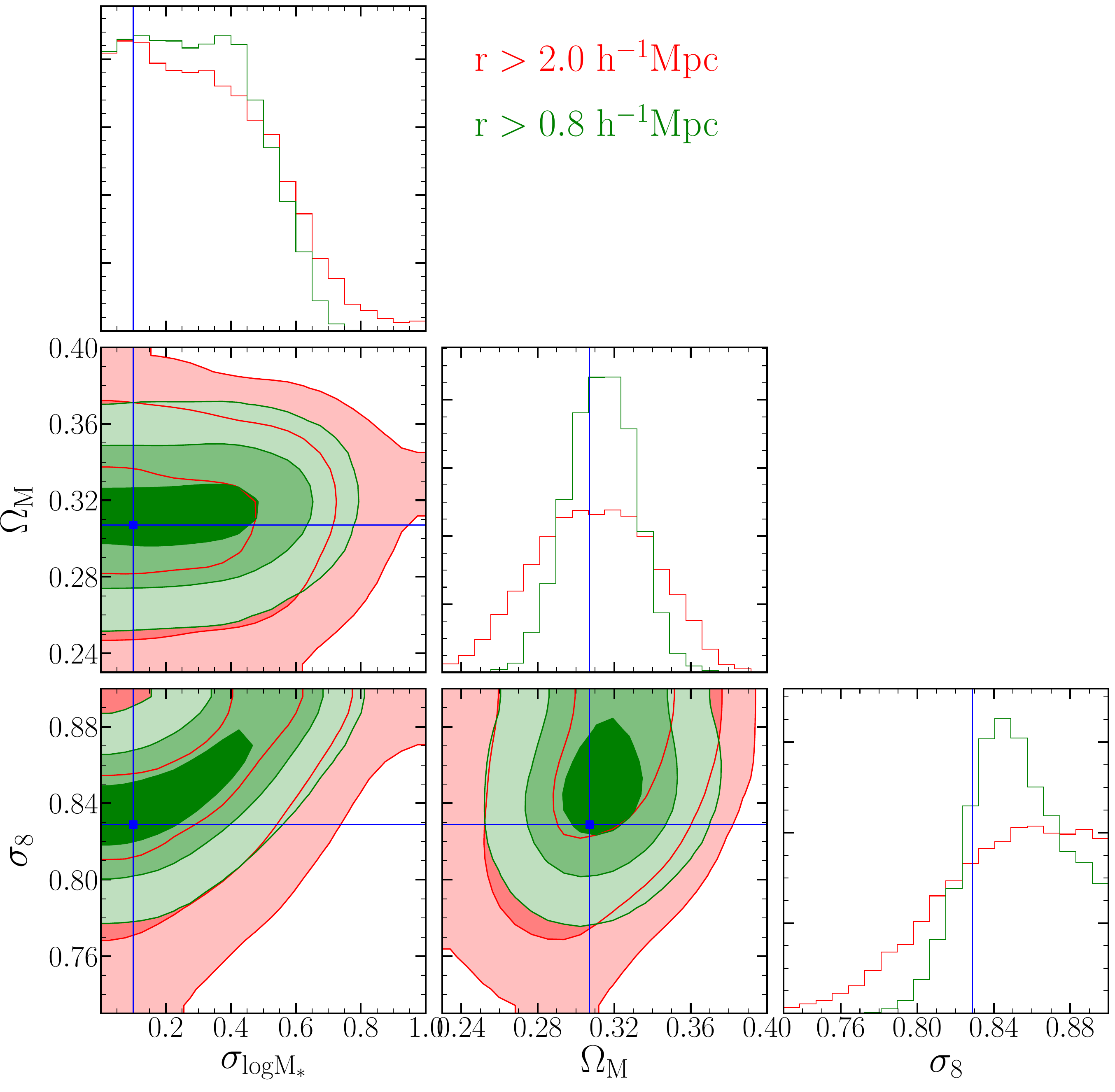}
\caption{Marginalized $1\sigma$, $2\sigma$, and $3\sigma$ credibility regions in $\Omega_{\rm m}$, $\sigma_8$ and $\sigma_{\log M_*}$ (the scatter of the SHAM) derived from the projected correlation function of a mock galaxy sample with $\bar{n} = 1.14 \times 10^{-2} h^3 {\rm Mpc^{-3}}$. The red contours show the constrains obtained using scales above $2\ h^{-1}{\rm Mpc}$ whereas the green contours show the constrains obtained using scales above $0.8\ h^{-1}{\rm Mpc}$. The blue squares and lines indicate the true values adopted in our mock galaxy catalogue.}
\label{fig:constr}
\end{figure*}

So far we have considered $z=0$. This is the worst case scenario for the scaling technique as it is where most non-linear structure exists and also where the growth histories differ the most (at high redshifts all cosmologies are essentially Einstein-de-Sitter.). On the other hand, future observational endeavours will mostly measure structure at $z \gtrsim 1$ as they focus on mapping ever larger cosmological volumes. Thus, we next present our results at $z=1$.

In Fig.~\ref{fig:full_scaling_z1} we show a plot analogous to Fig.~\ref{fig:full_scaling} but for measurements at $z=1$. We can see that, as expected, the accuracy of the scaling increases for most parameters. For the 5 standard $\Lambda$CDM parameters, the mass power spectrum is almost indistinguishable from that measured in the full simulations; for haloes and subhaloes, the accuracy is almost always less than 2\% and typically at the 1\% level. The only case where the uncertainty of the scaling increases is for the real-space mass power spectrum when changing $w_0$ and $w_a$. In this case, the error can reach 5\% at $k\sim5\ihMpc$. The origin of this (comparatively large) error is inaccuracies of the \cite{Ludlow:2016} model in capturing the effect of dynamical dark energy in the concentration-mass relation. Thus, possible future improvements in this regard should translate into more accurate scaling predictions.

\section{Using the scaling technique to constrain cosmology}
\label{sec:MCMC}

We will now provide an illustration of the performance of the rescaling technique for cosmological analyses of galaxy clustering. For this, we will constrain cosmological parameters using the projected correlation function of a mock galaxy sample.

Our mock data catalogue aims at mimicking a magnitude-selected volume-limited sample of $M_r < -19.5$ SDSS-like galaxies but in a volume 4 times larger. This sample has a number density of $1.14 \times 10^{-2} h^3 {\rm Mpc^{-3}}$ \citep{Guo:2015}. We build such catalogue using a subhalo abundance matching technique (SHAM) on the outputs of our Planck-13 simulation (c.f. \S\ref{sec:the_sims} and Table~1). We adopt $V_{\rm peak}$ as a proxy for stellar mass in SHAM and assume a log-scatter of $\sigma_{\log \rm M_*} = 0.1$. We recall that SHAM has been shown to reproduce accurately the clustering of galaxies in state-of-the-art hydrodynamical simulations \citep[e.g.][]{Chaves-Montero:2016}. Therefore, we expect this procedure to adequately capture the complexities as well as the available cosmological information in galaxy clustering. 

For simplicity, our summary statistic of choice is the redshift-space two-point projected correlation function, $w_p(r_p)$. We compute this quantity using {\tt corrfunc} -- a heavily optimised tree code \citep[][]{Corrfunc1,Corrfunc2} -- and integrating pair separations along the line of sight up to $40\,\hMpc$. 

We model the uncertainty in these measurements with the following covariance matrix:

\begin{equation}
\label{eq:covariance}
 C_{i,j} =  \frac{1}{4}C^{\rm SDSS}_{i,j}\bigg[ \frac{w_p(r_i)}{w^{\rm SDSS}_p(r_i)} \bigg]^2 + \left[0.05\, w_p(r_i)\right]^2\,\delta^D_{ij}
\end{equation}

\noindent where $C^{\rm SDSS}_{i,j}$ is the covariance matrix of our target SDSS-like sample, as estimated by \cite{Guo:2015}. Since our SHAM and the observed correlation functions might differ, we renormalize this covariance matrix with the ratio between the observed and our mock $w_p$. We further divide the whole covariance by a factor of 4, mimicking the precision of a future larger galaxy survey (such as those aimed at by DESI-BGS or TAIPAN). 

Finally, the second term in Eq.~\ref{eq:covariance} represents an estimate of the uncertainty in the theoretical modelling, which, in our case, we conservatively estimate as an uncorrelated $3\%$ (motivated by Fig.~\ref{fig:full_scaling}). We did not include the cosmic variance nor shot noise associated to our original rescaling simulations, since we expect them to be much smaller than the other two terms in Eq.~\ref{eq:covariance} owing to our ``fixed-and-paired'' initial conditions.

We model the mock correlation function by applying a SHAM technique to the 3 simulations in our optimal set. As in \S\ref{sec:accuracy}, we select the best simulations based on an ANN trained on the theoretical expectation for the error at $k=1\,\ihMpc$. 

We will consider a 3-dimensional parameter space $\bm{\pi} \equiv (\Omega_{\rm m}$, $\sigma_8$, $\sigma_{\log M_*} )$. The prior on these parameters were $\Omega_{\rm m} \in [0.23 - 0.4]$, $\sigma_8 \in [0.73 - 0.9]$, and $\sigma_{\log M_*} \in [0-1]$. We note that we initially considered a larger $7$-dimensional parameter space, however, the constraints on all the other parameters were rather weak. Thus, for simplicity, we restrict ourselves to this smaller space.

We will consider the whole range of scales robustly measured in observations and predicted by our model. Explicitly, we set $0.8\,\hMpc$ as the minimum scale, since below the scaling technique has not been properly tested; and set $20\,\hMpc$ as the maximum, as larger scales will be heavily affected by finite-box effects. For comparison, we will also consider another case where we set the minimum scale to $2\,\hMpc$.

We model the probability of observing a particular set of correlation function values, $w_{p}$, as a multivariate Gaussian. Thus, the likelihood of a set of $\bm{\pi}$ values given the data is:

\begin{equation}
-2\ \ln\,\mathcal{L}(\bm{\pi}) \propto (w_p - m(\bm{\pi}))^{T}\,C^{-1}\,(w_p - m(\bm{\pi}))
\end{equation}

\noindent where $C^{-1}$ is the inverse of the covariance matrix given by Eq.~\ref{eq:covariance}, $w_p$ is the projected correlation function as measured in our mock data, and $m(\bm{\pi})$ is our theoretical model for such quantity. 	

We obtain posterior distribution functions for our parameters using an iterative Gaussian emulation of the likelihood \citep{Pellejero-Ibanez:2020}. We note that this method requires approximately 100 times less model evaluations than a traditional Monte Carlo-Markov Chain algorithm. For each step of the iterative emulation, we compute the model $m(\bm{\pi})$ by scaling the appropriate simulation in our optimal suite, applying SHAM to our subhalo catalogues, and measuring the projected correlation function in exactly the same way as we did for our mock data.

The marginalized constraints on the free parameters of our model are shown in Fig.~\ref{fig:constr}. The red contours show the constraints obtained using scales above $2\ h^{-1}{\rm Mpc}$ whereas the green contours show the respective constraints obtained using scales above $0.8\ h^{-1}{\rm Mpc}$. First of all, we find that the best fit parameters are all statistically compatible with the true values in the mock (denoted by blue squares). This confirms the accuracy of our approach as a whole, and the feasibility of its use for cosmological analyses.

It is worth to note the tight constraints we obtain for all the three parameters, especially for when using small scales: $\sim 8\%$ for $\Omega_m$ and $\sim4\%$ for $\sigma_8$. This highlights the potentially large amount of information encoded in the small-scale galaxy clustering as well as the importance of a predictive model to extract it. In fact, there is only a small degeneracy between $\sigma_8$ and $\sigma_{M_*}$, which can, in principle, be broken by other clustering statistics such as the redshift space correlation function, higher order statistics, cluster counts, etc., or by independent measures of the scatter between SHAM quantities and observed galaxy properties. We will explore the full constraining power of galaxy clustering in a future work.
 
\section{Summary and Conclusions}
\label{sec:Conclusions}

%The scaling technique modifies a dark matter simulation with a given cosmology so it mimics the outputs of a simulation ran with a different cosmology. 
In this paper we have shown that cosmology-rescaling together with a suite of only 3 $N$-body simulations can provide highly accurate predictions for the clustering of dark matter, haloes, and subhaloes over a wide range of cosmologies, including massive neutrinos and dynamical dark energy. We illustrated how this opens the remarkable possibility of exploiting of the galaxy correlation function on small scales for cosmological analyses.

Below we summarize the main results of this work:

\begin{itemize}
\item Using an analytical prediction for the rescaling accuracy, we found the optimal number and cosmologies of simulations to be rescaled. A suite of only 3 simulations minimizes the computational cost and maximizes precision over a wide cosmological parameter space (Fig.~\ref{Fig:BestN} and Fig.~\ref{Fig:NPerf}).

\item We improved the original version of the scaling algorithm by including a ``concentration correction''. This displaces particles inside haloes according to the expected difference in halo concentration  (Fig.~\ref{Fig:Profile}). This substantially improves the accuracy of the algorithm on small scales (Fig.~\ref{Fig:PredErr}).

\item We quantified the accuracy of the scaling technique over the range: $\sigma_8 \in [0.73, 0.9]$, $\Omega_{\rm m} \in [0.23, 0.4]$, $\Omega_b \in [0.04, 0.06]$, $n_s \in [0.92, 1.01]$, $M_{\nu} \in [0, 0.4]$, $w_0 \in [-1.15, -0.85]$, and $w_a \in [-0.3, 0.3]$. We showed that the nonlinear power spectrum for dark matter, haloes, and subhaloes -- both in real and redshift space, and at $z=0$ and $1$ -- can be obtained with a precision better than $3\,\%$, even down to sub-Mpc scales (Figs.~\ref{fig:full_scaling} and\ref{fig:full_scaling_z1}).

\item As an illustration of the potential of our approach, we explore the constraining power of the projected correlation function on small scales. For this, we performed a Bayesian analysis to fit the correlation function of mock $M_r < -19.5$ SDSS-like galaxies. Using subhalo abundance matching over scales $r \in [0.8-20]\,\hMpc$ we recover tight and unbiased constraints on $\Omega_{\rm m}$, $\sigma_8$, and $\sigma_{\log M_*}$ (Fig.~\ref{fig:constr}).
\end{itemize}

Overall, we conclude that the scaling technique can be an extremely useful tool to predict non-linear structure in the universe. Combined with high-resolution original simulations, advanced galaxy modelling is possible. This translates into accurate and predictive models for the distribution of galaxies as a function of cosmological parameters, which, in turn, are expected to provide significant improvements in the cosmological constraints derived from galaxy clustering.

\section*{Acknowledgements}

The authors acknowledge the support of the ERC Starting Grant number 716151 (BACCO). SC acknowledges the support of the ``Juan de la Cierva Formaci\'on'' fellowship (FJCI-2017-33816). 
The authors thankfully acknowledge the computer resources at MareNostrum and the technical support provided by Barcelona Supercomputing Center (RES-AECT-2019-2-0012)".

\bibliography{Biblio}

\appendix

\section{Scaling all parameters}
\label{sec:A1}

We now test the performance of the scaling when varying more than one parameter at a time. 

We chose 20 target cosmologies from two Latin-hypercube spanning our parameter space, and carried out the corresponding $N$-body simulations. In the first Latin-hypercube we only vary the standard LCDM parameters (ie. $\sigma_8,\ \Omega_{\rm m},\ \Omega_b,\ n_s\ \&\ h)$, that are the one we use to define the optimal position of the main simulation to minimize the error on it. The cosmological parameter of these target cosmologies, as well as the cosmologies used to scale from, are shown in the first 20 rows of table~\ref{Table:A1}. The second Latin-hypercube, also with 10 different cosmologies, additionally vary $\rm M_\nu$, $\rm w_0$ \& $\rm w_a$, and the value of their cosmological parameters are shown in the last 10 rows of table~\ref{Table:A1}. 

\begin{figure}
\includegraphics[width=0.38\textwidth]{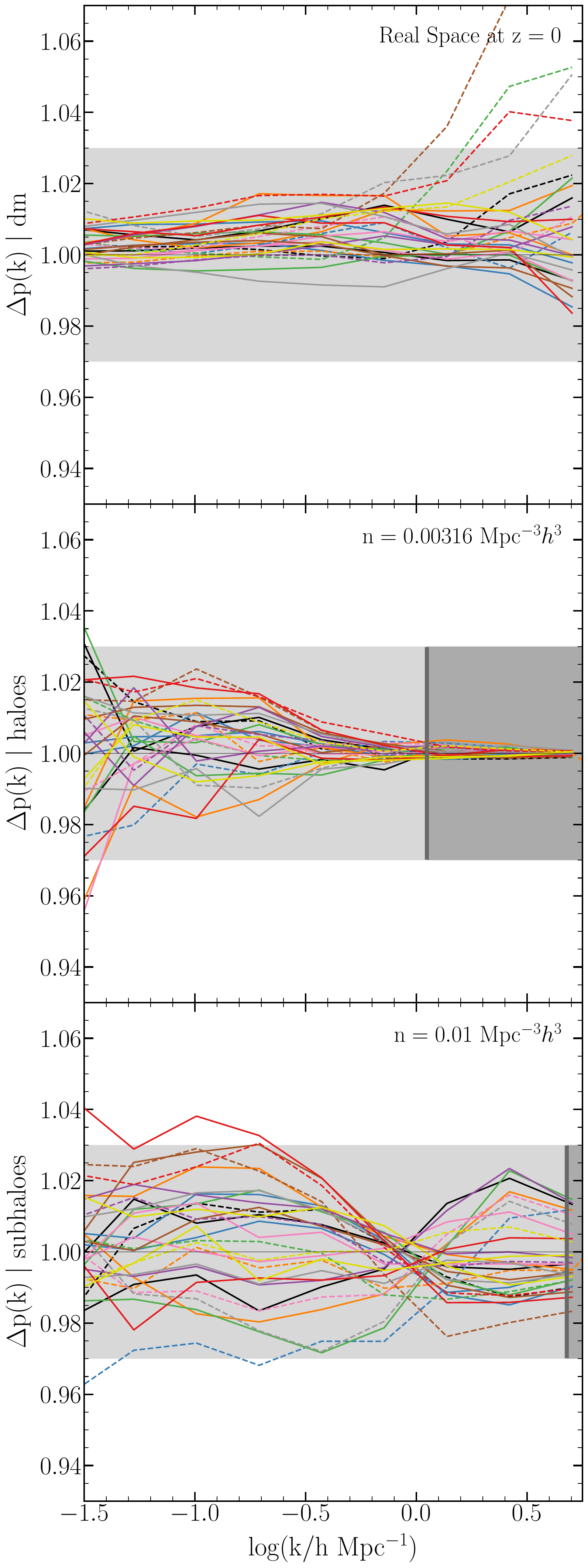}
\caption{The ratio between the power spectra of scaled and target simulations in real space at $z=0$. The target cosmologies are chosen using two Latin-hypercube over the cosmological space as explained in section~\ref{sec:A1}. 
The solid lines show the prediction of the scaling for cosmologies were we modify all the cosmological parameter used to optimize our model ($\sigma_8$, $\rm \Omega_{\rm m}$, $\rm \Omega_b$, $\rm n_s$ \& $\rm h$, first twenty cosmologies of table~\ref{Table:A1}), while the dashed lines show the prediction for cosmologies where we additionally change $\rm M_\nu$, $\rm w_0$ \& $\rm w_a$ (last ten cosmologies of table~\ref{Table:A1}).
The light shaded region highlights a $3\%$ discrepancy. The top panel displays results for dark matter, whereas middle and bottom panels do so for haloes with a number density of $\rm 3.16\times10^{-3}\,h^3Mpc^{-3}$ selected according to their mass, and subhaloes with a number density of $\rm 10^{-2}\ h^3Mpc^{-3}$ selected according to their peak maximum circular velocity ($\rm V_{peak}$). The dark shaded region shows the region where Poisson shot-noise is above than 80\% of the power spectrum amplitude.}
\label{fig:A1}
\end{figure}

The performance of the scaling at $z=0$ at real scale is shown in Fig.~\ref{fig:A1}. The top, middle and bottom panels show the ratio of the power spectrum between the target and scaled simulations of the dark matter particles, haloes and subhaloes, using the same selection as in the rest of the paper. The solid lines show the performance of the scaling on the simulation with cosmologies from our first Latin-hypercube, while the dashed lines are done with the second one. 

This test needed the run of 60 new simulations (30 cosmologies using pair simulation) and indicates that the performance of the scaling technique does not change when varying more than one parameter at a time. The expected error is typically below 2\% for the matter power spectrum, and 3\% for haloes and subhaloes. Only a handful of cosmologies show a deviation above  3\% on small scales and only for the mass power spectrum. This is because the scaling technique in this regime is dominated by the concentration correction. As explained in section~\ref{sec:Improvements}, for cosmologies with $\rm M_\nu$, $\rm w_0$ \& $\rm w_a$, there is no  accurate mass-concentration model available. 
%On a future release, we will expand the actual mass-concentration models for this kind of cosmologies, improving the performance of the scaling on this regime.

These results support the robustness of the scaling technique as well as indicating that 3\% is a fair estimate of the accuracy of this technique for the power spectrum. Although not shown here, we have repeated this test for target samples at $z=1$ and in redshift space, finding similar results.

\begin{table}
\caption{
The cosmological parameters chosen to test the scaling technique when varying all parameters at the same time. For the first twenty cosmologies, we only vary the parameters used to optimize the scaling ($\sigma_8$, $\rm \Omega_{\rm m}$, $\rm \Omega_b$, $\rm n_s$ \& $\rm h$, solid lines in Fig.~\ref{fig:A1}) whereas for the next ten cosmologies, we additionally change $\rm M_\nu$, $\rm w_0$ \& $\rm w_a$ (dashed lines on Fig.~\ref{fig:A1}). The first column indicates the cosmology of the simulation used to scale to the target cosmology.}
\begin{center}
 \begin{tabular}{c c c c c c c c c c}
 \hline
 $\rm  Original\ cosmology$ & $\sigma_8$ & $\rm \Omega_{\rm m}$ & $\rm \Omega_b$ & $\rm n_s$ & $\rm h$  & $\rm M_\nu$ & $\rm w_0$ & $\rm w_a$\\ 
 \hline
Narya & 0.864 & 0.342 & 0.049 & 0.990 & 0.796 & 0.000 &  0.000 & 0.000 \\ 
Nenya & 0.752 & 0.325 & 0.055 & 0.967 & 0.668 & 0.000 &  0.000 & 0.000 \\ 
Nenya & 0.779 & 0.374 & 0.050 & 0.929 & 0.695 & 0.000 &  0.000 & 0.000 \\ 
Vilya & 0.801 & 0.268 & 0.048 & 0.960 & 0.732 & 0.000 &  0.000 & 0.000 \\ 
Nenya & 0.868 & 0.300 & 0.043 & 1.002 & 0.710 & 0.000 &  0.000 & 0.000 \\ 
Narya & 0.835 & 0.393 & 0.046 & 0.981 & 0.643 & 0.000 &  0.000 & 0.000 \\ 
Nenya & 0.887 & 0.355 & 0.058 & 0.950 & 0.620 & 0.000 &  0.000 & 0.000 \\ 
Vilya & 0.785 & 0.242 & 0.053 & 0.997 & 0.620 & 0.000 &  0.000 & 0.000 \\ 
Nenya & 0.830 & 0.286 & 0.057 & 0.929 & 0.752 & 0.000 &  0.000 & 0.000 \\ 
Vilya & 0.744 & 0.256 & 0.041 & 0.944 & 0.768 & 0.000 &  0.000 & 0.000 \\  
Vilya & 0.796 & 0.284 & 0.053 & 0.957 & 0.667 & 0.000 &  0.000 & 0.000 \\
Narya & 0.884 & 0.380 & 0.057 & 0.992 & 0.638 & 0.000 &  0.000 & 0.000 \\
Vilya & 0.771 & 0.241 & 0.042 & 1.006 & 0.719 & 0.000 &  0.000 & 0.000 \\
Narya & 0.762 & 0.360 & 0.051 & 0.978 & 0.784 & 0.000 &  0.000 & 0.000 \\
Nenya & 0.851 & 0.309 & 0.047 & 0.973 & 0.722 & 0.000 &  0.000 & 0.000 \\
Nenya & 0.876 & 0.267 & 0.041 & 0.948 & 0.769 & 0.000 &  0.000 & 0.000 \\
Nenya & 0.811 & 0.347 & 0.059 & 0.988 & 0.749 & 0.000 &  0.000 & 0.000 \\
Narya & 0.823 & 0.400 & 0.049 & 0.945 & 0.693 & 0.000 &  0.000 & 0.000 \\
Vilya & 0.732 & 0.319 & 0.056 & 0.937 & 0.658 & 0.000 &  0.000 & 0.000 \\
Vilya & 0.835 & 0.254 & 0.045 & 0.928 & 0.617 & 0.000 &  0.000 & 0.000 \\
Narya & 0.758 & 0.355 & 0.042 & 1.003 & 0.742 & 0.060 & -0.898 & 0.058 \\
Vilya & 0.774 & 0.269 & 0.055 & 0.957 & 0.787 & 0.130 & -1.119 & -0.054 \\
Nenya & 0.741 & 0.393 & 0.056 & 0.983 & 0.622 & 0.228 & -1.041 & -0.125 \\
Vilya & 0.888 & 0.241 & 0.059 & 0.934 & 0.718 & 0.168 & -0.913 & 0.140 \\
Vilya & 0.781 & 0.297 & 0.052 & 0.940 & 0.689 & 0.076 & -1.002 & -0.263 \\
Nenya & 0.860 & 0.300 & 0.048 & 0.954 & 0.618 & 0.312 & -0.854 & 0.094 \\
Nenya & 0.803 & 0.330 & 0.050 & 0.969 & 0.735 & 0.183 & -0.948 & -0.108 \\
Vilya & 0.828 & 0.254 & 0.048 & 0.998 & 0.660 & 0.393 & -1.087 & -0.210 \\
Narya & 0.841 & 0.369 & 0.045 & 0.928 & 0.779 & 0.331 & -0.989 & 0.212 \\
Nenya & 0.880 & 0.332 & 0.044 & 0.979 & 0.666 & 0.269 & -1.135 & 0.273 \\
\end{tabular}
\end{center}
\label{Table:A1}
\end{table}

% Don't change these lines
\bsp	% typesetting comment
\label{lastpage}
\end{document}